\documentclass[12pt]{iopart}
\usepackage{iopams}  
\usepackage{graphicx}
\usepackage{color}
\newcommand{\myPr}{{\rm Pr}}
\newcommand{\myVec}[1]{{\underline{#1}}}

\begin{document}

\title[Numerical entropy estimation and complexity -- $2D$ Ising Ferromagnet]{A computational mechanics approach to estimate entropy and (approximate) 
complexity for the dynamics of the $2D$ Ising Ferromagnet}

\author{O Melchert and A K Hartmann}
\address{Institut f\"ur Physik, 
Universit\"at Oldenburg, 
Carl-von-Ossietzky Strasse, 
26111 Oldenburg, Germany}
\ead{melchert@theorie.physik.uni-oldenburg.de and alexander.hartmann@uni-oldenburg.de}

\begin{abstract}
We present a numerical analysis of the entropy rate and statistical complexity related to the spin flip dynamics
of the $2D$ Ising Ferromagnet at different temperatures $T$. 
We follow an information theoretic approach and test
three different entropy estimation algorithms to asses entropy rate and statistical complexity 
of binary sequences. The latter are obtained by monitoring the orientation of 
a single spin on a square lattice of side-length $L=256$ at a given temperature parameter over time.
The different entropy estimation procedures are based on the $M$-block
Shannon entropy (a well established method that yields results for benchmarking purposes), 
non-sequential recursive pair substitution (providing an elaborate and an approximate estimator) 
and a convenient data compression algorithm
contained in the {\tt zlib}-library (providing an approximate estimator only). 
We propose an approximate measure of statistical 
complexity that emphasizes on correlations within the sequence and which is easy to implement, even
by means of black-box data compression algorithms. 
Regarding the $2D$ Ising Ferromagnet simulated 
using Metropolis dynamics and for binary sequences
of finite length, the proposed approximate complexity measure is peaked close to the 
critical temperature. For the approximate estimators, a finite-size scaling 
analysis reveals that the peak approaches the critical temperature as the 
sequence length increases.
Results obtained using different spin-flip dynamics are briefly discussed.
The suggested complexity measure can be extended to non-binary 
sequences in a straightforward manner.
\end{abstract}

\maketitle

\section{Introduction}

The basic task of data compression algorithms is to discover \emph{patterns}
(synonymous with regularities, correlations, symmetries and structure; see 
Sect.\ II of 
Ref.\ \cite{Shalizi2001}), and to remove the respective redundancies 
from supplied input data in order to minimize the space required to store the data.
Interestingly, the pattern discovery and data compression process of particular data
compression schemes finds application in contexts as diverse as
as e.g.\ DNA sequence classification \cite{Loewenstern1995}, 
entropy estimation \cite{Ebeling1997,Baronchelli2005,Grassberger2002}, and,
more generally, time series analysis \cite{Puglisi2003}.

Correspondingly, in the analysis of complex systems, one wants to
find a measure for the information-theoretic
``complexity'' of a system \cite{crutchfield2010}. 
The most simple
measure is the entropy \cite{cover2006}, but this is maximal
 for a purely random system.
This contradicts the basic idea of complexity, which involves
some structure, but not just regular structure. On the other hand, other
measures of complexity, like a (minimal) algorithm/computer/circuit
 able of generating
(an instance of) the problem \cite{kolmogorov1963,chaitin1987,machta2006}, 
 are often impractical when it
comes to the analysis of given large systems. Hence,
data compression algorithms are a natural and in particular simple 
candidate for detecting complexity \cite{Baronchelli2005,Nagaraj2011}. 
Note that also other
practically computable approaches exist, which indeed seem
to measure complexity as expected, like mutual information 
\cite{grassberger1986,cover2006} or statistical complexity 
\cite{crutchfield1989}.

 As a first step, different proposed quantities 
 are usually applied to simple toy systems, e.g.\
models exhibiting only few states
\cite{crutchfield1989,Baronchelli2005,wiesner2011,Nagaraj2011}. 
In statistical mechanics on the other hand, 
one studies models which involve many
degrees of freedom with non-trivial interactions. Such models
are regarded as being 
very complex often right at phase transitions \cite{goldenfeld1992}.
 This large degree of complexity is from the physical point of view
 visible via  growing correlations in the system.
An information-theoretic analysis of the complexity of such models
has to our knowledge been considered so far only in few studies
and only by example \cite{Arnold1996}.
 The aim of this work 
is to study extensively data compression algorithms for time series
generated by the Ising model, which is one of the most fundamental 
and important models
of statistical mechanics exhibiting a phase transition. In particular,
we want to find out whether the phase transition can be detected, located
and analyzed numerically with high precision just by looking at
complexity measures derived from symbol substitution
methods which can be used for data compression.

By means of these methods, we attempt to estimate the 
\emph{entropy rate} of symbolic 
sequences and we compute a measure to account for correlations that possibly characterize the sequence.
The latter observable, here referred to as \emph{approximate complexity}, is related to the 
excess entropy which characterizes the \emph{statistical complexity} of the sequence and
it can very well be understood in 
terms of information theory. Further, it can easily be computed 
by means of black-box data compression algorithms, as, e.g., the {\tt compress}
algorithm contained in the {\tt zlib}-library \cite{zlib}.
While the entropy accounts for the randomness contained in the sequence, the approximate
complexity is sensitive to correlations.

The symbol substitution method considered in the bulk of the presented article 
is a particular dictionary based data compression scheme that operates by a
\emph{non-sequential recursive pair substitution} process and is hence referred
to as NSRPS. 
The basic routine of the NSRPS algorithm is a \emph{pair 
substitution} step that amounts to replace the most frequent 
two-symbol-subsequence by a new symbol that is shorter in length. 
If this process is performed in repeated manner, it is possible to achieve 
a compression of the input data. 
Such pair substitution methods, intending to quantify the degree of ``patterness'' of symbolic
sequences, 
where introduced several decades ago by Ebeling and Jim\'enez-Monta\~no, 
see Ref.\ \cite{Ebeling1980}. In Ref.\ \cite{JimenezMontano2002}, a sequence compressing
algorithm based on the NSRPS paradigm was 
introduced and used to estimate the information content of binary sequences. 
Rigorous results on the NSRPS method, presented by Benedetto, Caglioti, and Gabrielle in 
Ref.\ \cite{Benedetto2006}, imply that for sufficiently large sequences the NSRPS method can
be utilized to estimate the entropy rate of an ergodic process.
On this basis, Calcagnile, Galatolo, and Menconi recently reported on numerical experiments,
see Ref.\ \cite{Calcagnile2010}, that compared entropy estimates arising from the NSRPS 
method and other well established methods considering different maps and a stationary process
known as ``renewal process''. The authors found that the NSRPS method provided the best
approximation to the respective entropy values.

Most recently, a NSRPS based randomness measure 
for symbolic sequences was introduced and tested for short 
sequences \cite{Nagaraj2011}. 
In the latter study, the sequences where obtained from iterating the logistic map at different 
bifurcation parameters and applying a proper discretization procedure. 
Therein, the NSRPS based randomness measure appeared to be strongly 
correlated to the Lyapunov exponent of the map and hence could be 
used to quantify whether a particular sequence appears to have a simple or 
a random structure (note that the authors of Ref.\ \cite{Calcagnile2010} refer to it as 
``complexity measure''. More precise, this type of complexity is termed \emph{deterministic
complexity} \cite{Feldman2008} and it merely measures the randomness associated
to a symbolic sequence. As regards this notion of deterministic complexity, 
the randomness measure presented in \cite{Nagaraj2011} is simply proportional to 
the entropy rate.).
Albeit the NSRPS method is of academic interest (as documented by the references above), 
it is commonly not used in practice. More frequently used dictionary 
compression algorithms are based on Lempel-Ziv (LZ) coding \cite{Ziv1977}.
E.g., the {\tt compress} data-compression tool available in the {\tt zlib}-library uses LZ77 compression, 
a particular variant of LZ coding. 
Effectively, using LZ coding, subsequences of the input sequence are replaced by
pointers that specify positions within the sequence where the respective  
subsequences have occurred earlier \cite{Ziv1977,Bell1989}. This latter approach to compression is slightly 
different from the iterated pair substitution on which the NSRPS method
is build upon.
The entropy rate and approximate complexity of a sequence can also be approximated by means of the NSRPS based 
randomness measure and by using the {\tt compress}-algorithm. Both methods, however, 
allow only to compute an upper bound and are not as precise as the more elaborate
estimates.

In this work, 
we aim to assess how well the NSRPS based entropy estimation method
described by Ref.\ \cite{Calcagnile2010} performs on binary sequences that represent the 
spin-flip dynamics of the $2D$ Ising Ferromagnet (FM) at different temperatures $T$. For this purpose
we consider single-spin-flip Metropolis dynamics (main part of the presented article) 
as well as spin-flip dynamics induced by the Wolff cluster algorithm (results reported in 
subsection \ref{ssect:results_wolffClusterAlg}).
The input data to be analyzed is given 
by sequences $\myVec{S}=(s_1,\ldots,s_N)$ of length $N$, consisting of symbols 
$s_i$ over the binary alphabet $\mathcal{A}=\{0,1\}$. These sequences are obtained by monitoring
the time-series related to the orientation of a \emph{single} spin, 
located on a square lattice of side length $L$
with fully periodic boundary conditions.
In order to allow for a comparison of the results, we also estimate the entropy rate 
and complexity of the binary sequences by a well established approach based on the Shannon entropy, 
as presented in Ref.\ \cite{Crutchfield2003}. Previously, the latter approach led to the analysis of 
complexity-entropy diagrams that allow for a characterization of the temporal and spatial dynamics of
various stochastic processes, including simple maps as well as
Ising spin-systems, in purely information-theoretic coordinates \cite{Feldman2008}. 
We find that for the whole range of temperatures considered, the entropy rates and approximate
complexities estimated 
via the elaborate NSRPS algorithm of Ref.\ \cite{Calcagnile2010} are in good agreement with 
the Shannon entropy based estimates following Ref.\ \cite{Crutchfield2003}.
Furthermore, we find that the approximate complexity is peaked at a sequence-length dependent, effective critical temperature.
A finite-size scaling analysis in the sequence length (where the size of the Ising model that supplies the binary input sequences is 
fixed to $256\times 256$ spins) 
reveals that in the limit of infinitely long sequences, the peak is located close by the critical
temperature $T_c\approx 2.269$ of the $2D$ Ising Ferromagnet.

The remainder of the article is organized as follows. In Sect.\ 
\ref{sect:patternDiscovery} we introduce the 
well-established
information theoretic 
notation, the entropy rate and the approximate
complexity. For illustration and comparison, we also include
the results of these
observables for the $2D$ Ising FM as function of temperature.
In Sect.\ \ref{sect:symbSubs} we discuss the symbolic substitution 
method and the 
three different entropy estimation procedures. The
main part of this work are the results for theses procedures
and an assess of the 
performance, as presented in Sect.\ \ref{sect:results}. Finally, 
Sect.\ \ref{sect:summary} concludes with a brief summary.
A more elaborate summary of the presented article is available at the {\emph{papercore database}}
\cite{papercore}.

\section{Basic notation from information theory and pattern discovery}
\label{sect:patternDiscovery} 
%{{{1 
In subsection \ref{ssect:notation} we introduce basic notation from information theory,
needed to motivate the observables that are considered in the remainder of 
the article. First and foremost, these are the entropy rate and excess entropy that might
be associated to a sequence of symbols. 
If not stated explicitly, we thereby follow the notation used 
by Shalizi and Crutchfield \cite{Shalizi2001}, and Crutchfield and Feldman \cite{Crutchfield2003}.
In subsection \ref{ssect:entropyRate_complexity_IsingFM} we then 
discuss entropy rate and excess entropy as well as the convergence 
properties of the entropy rate by considering 
data obtained for the $2D$ Ising Ferromagnet at different temperatures 
including the critical point. Note that the results for the 
entropy rate and excess entropy
are similar to those presented by Arnold in Ref.\ \cite{Arnold1996}, 
hence, they are included here for illustration and comparison only.
On the other hand,
the convergence of the 
entropy rate, which in turn leads to an easy to compute approximate 
measure of  complexity, has to our knowledge not been discussed yet. 
Finally, in subsection \ref{ssect:approximateComplexity} we motivate 
an ``approximate complexity'' that quantifies for which parameters a 
given model exhibits a small or large statistical complexity.

\subsection{Block entropy, entropy rate convergence and complexity}
\label{ssect:notation}

%%%%%%%%%%%%%%%%%%%%%%%%%%%%%%%%%%%%%%%%%%%%%%%%%%%%%%%%
% FIGURE:       example_figs.eps
% LABEL :       fig:example:figs
%%%%%%%%%%%%%%%%%%%%%%%%%%%%%%%%%%%%%%%%%%%%%%%%%%%%%%%%
%{{{2
\begin{figure}[t!]
\begin{center}
\includegraphics[width=0.96\linewidth]{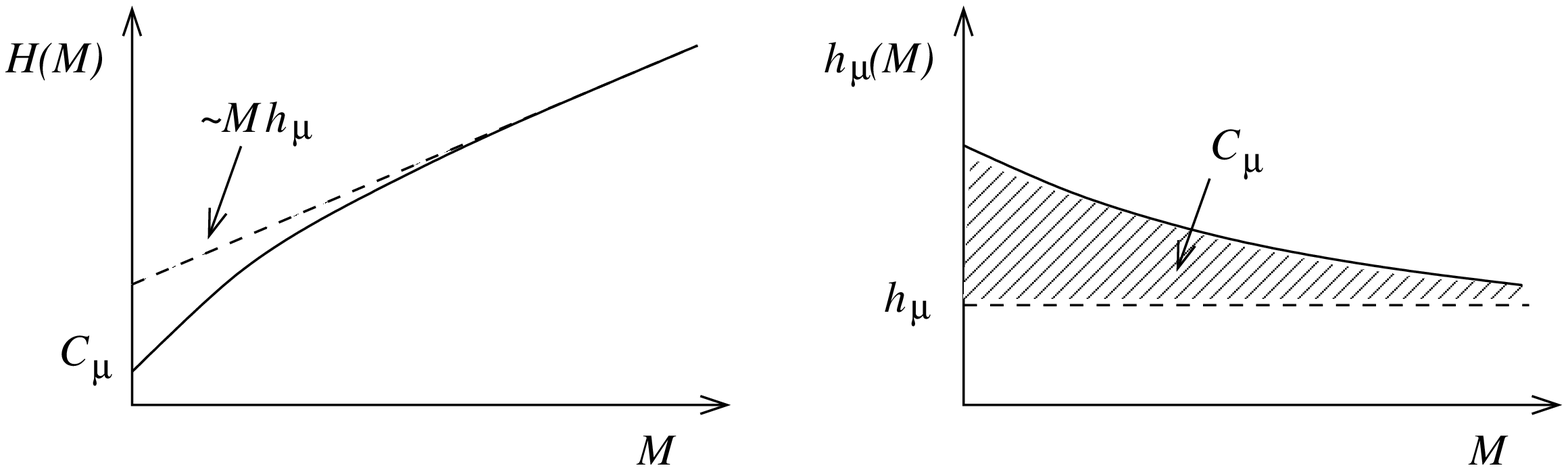}
\end{center}
\caption{
\label{fig:example_figs}
Schematic plots of the block entropy   $H(M)$ (left) and the
apparent entropy rate (or entropy gain) 
$h_\mu(M) = H(M)-H(M-1)$ (right), indicating
the relationship to the asymptotic entropy rate
$h_\mu=\lim_{M\to \infty}h_\mu(M)$ and the (statistical) complexity
$C_\mu= \sum_{M=1}^{\infty} [h_\mu(M)-h_\mu]$.
}
\end{figure}
%}}}2%%%%%%%%%%%%%%%%%%%%%%%%%%%%%%%%%%%%%%%%%%%%%%%%%%%

As pointed out in the introduction, the presented article addresses two issues: 
numerical estimation of the \emph{entropy rate} 
associated to a sequence of symbols, 
and providing an approximate measure of \emph{statistical complexity} that effectively 
accounts for the rapidity of entropy convergence.
A prerequisite needed to define the subsequent observables is the Shannon entropy
related to blocks of $M$ consecutive variables in a length $N$ sequence $\myVec{S}=(s_1,s_2,\ldots,s_N)$. 
This quantity is defined as 
\begin{eqnarray}
H(M) \equiv - \sum_{s^M \in \mathcal{A}^M} 
\myPr(s^M) \log_2[\myPr(s^M)], \label{eq:mBlockSE}
\end{eqnarray}
and is henceforth referred to as $M$-block Shannon entropy
the sum runs over all possible strings $s^M$ of $M$ symbols ($M\geq 1$)
from the alphabet $\mathcal{A}$,
and where $\myPr(s^M)$ denotes the probability (i.e.\ the empirical rate
of occurrence) of $s^M$ in the given sequence (where we agree to set $0 \log_2(0) \equiv 0$). 
For a schematic plot of $H(M)$, see Fig.\ \ref{fig:example_figs}.
In general, $H(M)$ is a nondecreasing function of $M$ bounded 
by $H(M)\leq M\cdot \log_2{|\mathcal{A}|}$, which is obtained if
the probability of a string factorizes and each letter has the
same probability of occurrence, i.e., $\myPr(s^M)=1/|\mathcal{A}|^M$.
In the limit of large block-sizes, $H(M)$ might not converge to a finite
value. As a remedy, due to the above bounding value, the \emph{entropy rate} $h_\mu$ 
is considered instead.
The entropy rate (also termed \emph{per-symbol entropy}) 
specifies the asymptotic rate of 
increase of the $M$-block Shannon entropy regarding the block length, i.e.\
\begin{eqnarray}
h_\mu \equiv \lim_{M\to \infty} \frac{1}{M} H(M). \label{eq:entropyRate}
\end{eqnarray}
For a given sequence it quantifies the randomness that remains after patterns 
on subsequences of increasing length are taken into account. 
In order to quantify its convergence properties, it is useful to consider
finite-$M$ approximations to the entropy rate.
By considering the \emph{entropy gain} $\Delta H(M)=H(M)-H(M-1)$ 
(with $H(0)\equiv 0$) it is
possible to show that $h_\mu=\lim_{M\to \infty}\Delta H(M)$ 
(see Sect.\ III\,B of Ref.\ \cite{Crutchfield2003}). Thus, one
also uses the term  \emph{apparent entropy rate}
\begin{eqnarray}
h_\mu(M) \equiv \Delta H(M) = H(M)-H(M-1). \label{eq:apparentEntropyRate1}
\end{eqnarray}
Alternatively one might refer to Eq.\ (\ref{eq:entropyRate}) and define 
\begin{eqnarray}
h^\prime_\mu(M) \equiv H(M)/M , \label{eq:apparentEntropyRate2}
\end{eqnarray}
where both definitions 
(Eqs.\ (\ref{eq:apparentEntropyRate1}) and (\ref{eq:apparentEntropyRate2})) 
are restricted to $M>0$.
Asymptotically it holds that $\lim_{M\to \infty} h^\prime_\mu(M)=\lim_{M\to \infty} h_\mu(M)$,
but $h^\prime_\mu(M)$ typically converges 
slower than $h_\mu(M)$, see Ref.\ \cite{Schuermann1996}.
A measure that quantifies how much the entropy rate at block length $M$ exceeds the 
actual entropy rate is given by the per-symbol $M$ redundancy
\begin{eqnarray}
r(M)\equiv h_\mu(M)-h_\mu. \label{eq:Mredundancy}
\end{eqnarray}
A value $r(M)> 0$ indicates that upon considering blocks of length $M$, the asymptotic entropy
rate of the sequence is overestimated. This overestimation is due to redundant information, i.e.\
 patterns, that characterizes the sequence. 
Summing up all per-symbol $M$ redundancies yields the \emph{excess Entropy}, which might 
also be referred to as (statistical) \emph{complexity} $C_\mu$ (see Eqs.\ 
(2)--(4) of Ref.\ \cite{Arnold1996}): 
\begin{eqnarray}
C_\mu\equiv \sum_{M=1}^{\infty} r(M) = \sum_{M=1}^{\infty} [h_\mu(M)-h_\mu]. \label{eq:complexity}
\end{eqnarray}
In order to compare Eq.\ (\ref{eq:complexity}) above to Eqs.\ (3) and (4) of 
Ref.\ \cite{Arnold1996}, note that the sum in 
the former equation can be interpreted as an integration of the discrete function $h_\mu(M)-h_\mu=\Delta H(M) - h_\mu$. 
Now, bearing in mind that the entropy gain $\Delta H(M)$ signifies the discrete derivative 
of the entropy itself, directly leads to 
\begin{eqnarray}
C_\mu=\lim_{M\to\infty}[H(M)-M h_\mu]. \label{eq:complexity2}
\end{eqnarray}
For large blocksize this implies the scaling $H(M)\sim C_\mu+M h_\mu$, 
allowing for a geometric interpretation of entropy rate and complexity: 
the asymptotic straight line approximation to $H(M)$ gives rise to 
the randomness, while the intercept with the 
$y$-axis is equal to the complexity, see right of Fig.\ \ref{fig:example_figs}.

%%%%%%%%%%%%%%%%%%%%%%%%%%%%%%%%%%%%%%%%%%%%%%%%%%%%%%%%
% FIGURE:       entropy/redundancy convergence function of blocksize
% LABEL :       fig:entropy_redundancy_convergence
%%%%%%%%%%%%%%%%%%%%%%%%%%%%%%%%%%%%%%%%%%%%%%%%%%%%%%%%
%{{{2
\begin{figure}[t!]
\begin{center}
\includegraphics[width=0.48\linewidth]{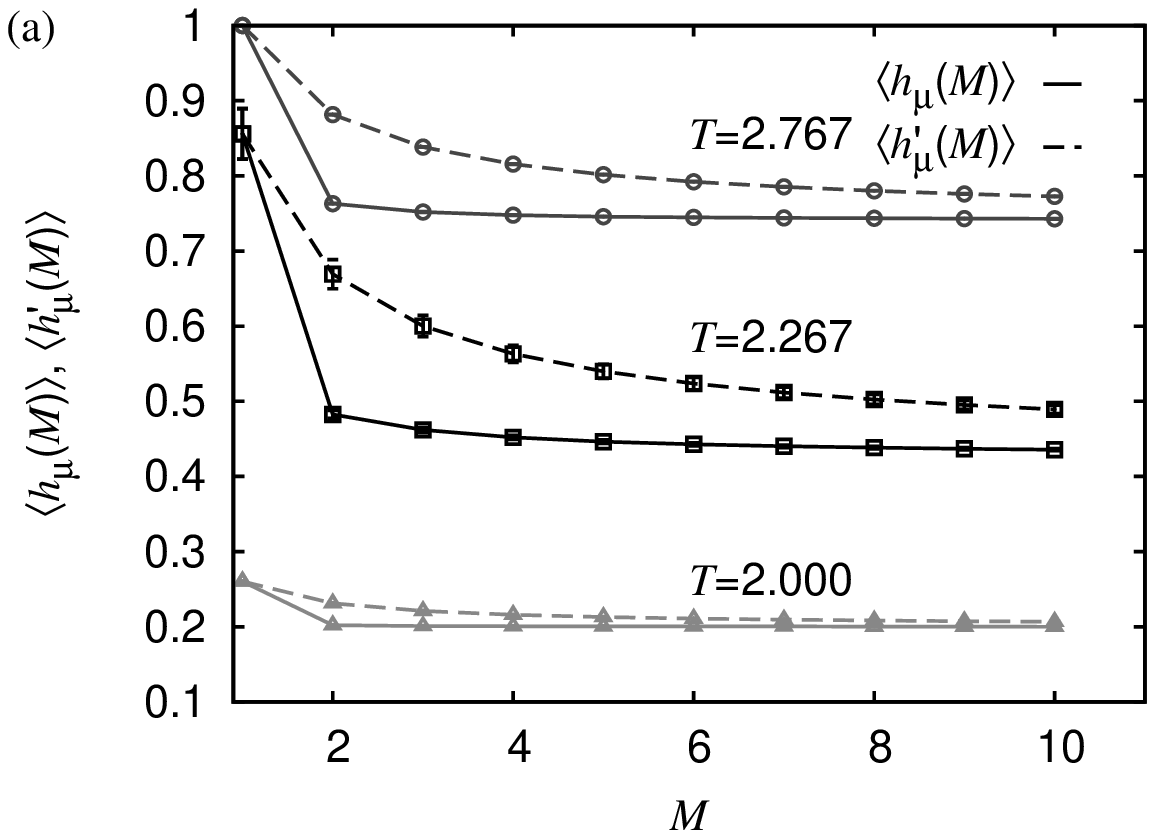}
\includegraphics[width=0.48\linewidth]{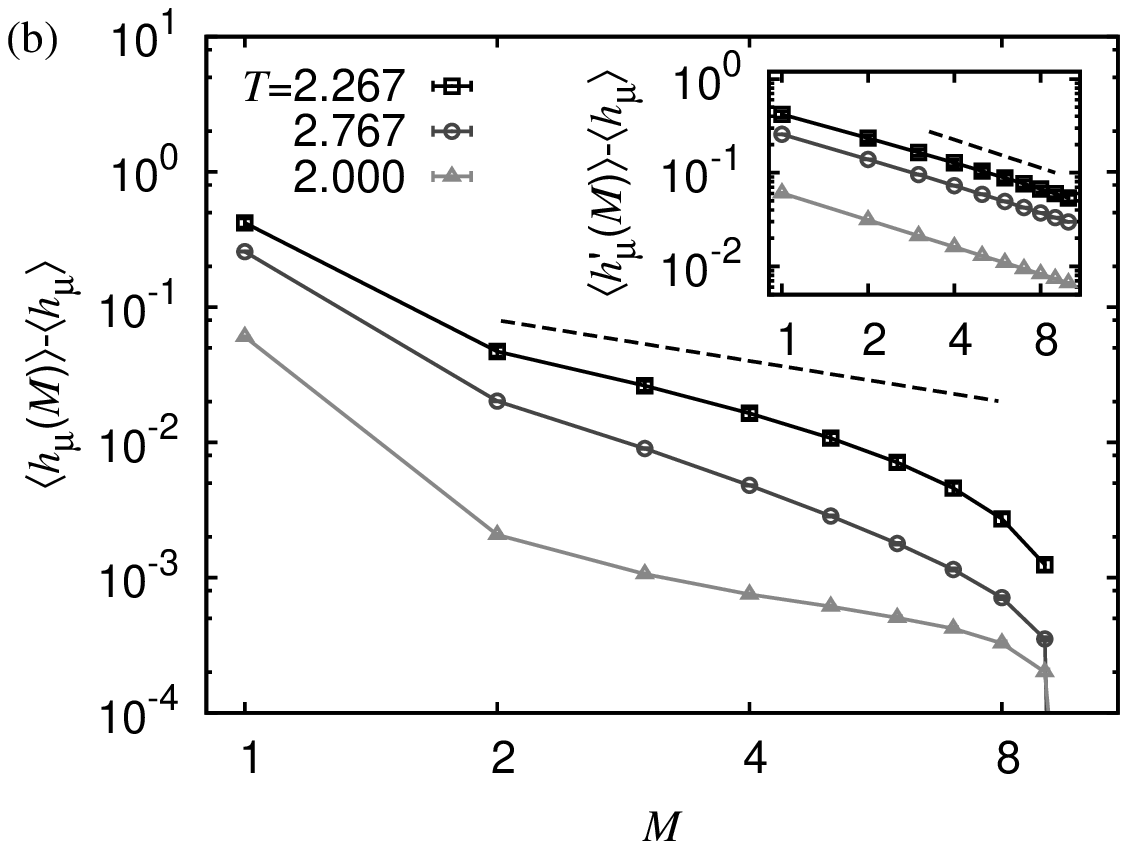}
\end{center}
\caption{
\label{fig:entropy_redundancy_convergence}
Scaling properties of the apparent entropy rates for three 
exemplary temperatures located below, close to, and above 
the critical temperature. For all quantities an average
$\langle \ldots \rangle$ over independent sequences of length $N=10^6$ 
is computed.
(a) the estimator $h_\mu(M)$ for the apparent entropy 
rate converges much faster than the estimator $h_\mu^\prime(M)$. 
(b) the associated per-symbol $M$ redundancies $h_\mu(M)-h_\mu$
decay faster than $\propto 1/M$ (for $M$ large enough this holds
also for $h_\mu^\prime(M)-h_\mu$, see inset). The dashed line
indicates a scaling of $\propto 1/M$.
}
\end{figure}
%}}}2%%%%%%%%%%%%%%%%%%%%%%%%%%%%%%%%%%%%%%%%%%%%%%%%%%%

\subsection{Results for the $2D$ Ising Ferromagnet}
\label{ssect:entropyRate_complexity_IsingFM}

As pointed out in the introduction, the sequences that are under 
scrutiny here are 
obtained by simulating a $2D$ Ising FM on a regular square lattice
with side-length $L$, using Metropolis dynamics \cite{newman1999}
at a selected temperature 
$T\in [2,\,2.8]$. 
For the purpose of modeling the binary sequences, a particular spin on the 
lattice is chosen as a ``source'', emitting symbols from the binary alphabet
$\mathcal{A}=\{0,1\}$ (after a simple transformation of the spin variables). 
Therefore, the orientation of the source-spin is
monitored during a number of $N$ Monte Carlo (MC) sweeps to yield a 
particular length $N$ sequence. 
Before the spin orientation is recorded, a sufficient number 
of sweeps are performed to ensure that the system is equilibrated.
In this regard, for a square lattice with side length $L=128$, and by analyzing the magnetization
of the system, we observed an equilibration time of approximately 
$\tau_{\rm eq}=3000$ MC 
sweeps for the lowest temperature. 
However, for each system considered we discarded the first $10^5$ sweeps to avoid
initial transients.

In the numerical experiments, the $M$-block Shannon entropy 
can only be computed for block-sizes smaller than some maximal size $M_{\rm max}$. 
Otherwise, for an input sequence of finite length, the (true) distribution of possible configurations
associated to blocks of $M$ consecutive variables will be approximated poorly (see Refs.\ \cite{Schuermann1996,Weiss2000}).
Effectively, this provides only an upper bound which might nevertheless yield
a reasonable approximation to the actual entropy of the considered sequences.
Further, the sum in Eq.\ (\ref{eq:complexity}) needs to be truncated 
to a finite number of terms, implying that only a lower bound on the excess entropy can be 
computed.
The quality of the lower bound is due to the rapidity of the convergence of the apparent 
entropy rate $h_\mu(M)$. 
In this regard, Fig.\ \ref{fig:entropy_redundancy_convergence}(a) shows the convergence for
the two estimators $h_\mu(M)$ and $h_\mu^\prime(M)$ of the apparent entropy rate as function 
of the block length $M$ for three exemplary temperatures located below, close to, and above the 
critical temperature. The input sequences had a length of $N=10^6$. 
As evident from the figure, $\langle h_\mu(M) \rangle$ converges substantially faster 
than $\langle h_\mu^\prime(M)\rangle$. The brackets 
$\langle\ldots\rangle$ indicate an average
over independent sequences. For comparison, at $T=2.267$ (i.e.\ close to the critical point) 
a fit to the functional form $\langle h_\mu^\prime(M)\rangle=h_\mu^\prime + a/M$ for the block-size interval $M\in[5,10]$ yields the 
parameters $h_\mu^\prime=0.4388(6)$ and $a=0.290(1)$, the reduced chi-square being $\chi^2/{\rm dof}=0.09$.
Note that the fit function above describes the data quite well, however, below we argue that 
the per-symbol $M$ redundancies decay somewhat faster than $\propto 1/M$ as the choice
of the particular scaling function suggests.
For comparison, at $M=10$ we find $\langle h_\mu(10)\rangle=0.436(1)$. For temperatures away from the critical 
point the estimates $\langle h_\mu^\prime \rangle$ and $\langle h_\mu(10) \rangle$ agree even better. 
Hence, in order to estimate the entropy rate $h_\mu$ and complexity $C_\mu$ we here consider the maximally 
feasible blocksize to be $M_{\rm max}=10$. Consequently, the (average) asymptotic entropy rate is set to 
$\langle h_\mu \rangle=\langle h_\mu(10) \rangle$. 
Again, note that this only provides an upper bound to the true entropy rate. The difference to the 
latter is due to long-range correlations in the sequences that are missed
by restricting the analysis to blocks of maximal length $M_{\rm max}=10$.
Fig.\ \ref{fig:entropy_redundancy_convergence}(b) shows 
the scaling properties of the 
per-symbol $M$ redundancies $\langle r(M)\rangle$ for the particular choice $M_{\rm max}=10$, defined in Eq.\ (\ref{eq:Mredundancy}).
As evident from the main plot of the figure, the redundancy $\langle h_\mu(M) \rangle-\langle h_\mu \rangle$ 
decays faster than $\propto 1/M$. 
For large enough block-size this holds also for the alternative definition $\langle h_\mu^\prime(M)\rangle-\langle h_\mu\rangle$, 
see inset of Fig.\ \ref{fig:entropy_redundancy_convergence}(b).
Such a scaling behavior is characteristic for \emph{finitary processes}, i.e.\ processes with 
a finite complexity $C_\mu$.

%%%%%%%%%%%%%%%%%%%%%%%%%%%%%%%%%%%%%%%%%%%%%%%%%%%%%%%%
% FIGURE:       apparent entropy rate/redundancy funtion of temperature
% LABEL :       fig:entropy_redundancy_convergence_temperature
%%%%%%%%%%%%%%%%%%%%%%%%%%%%%%%%%%%%%%%%%%%%%%%%%%%%%%%%
%{{{2
\begin{figure}[t!]
\begin{center}
\includegraphics[width=0.48\linewidth]{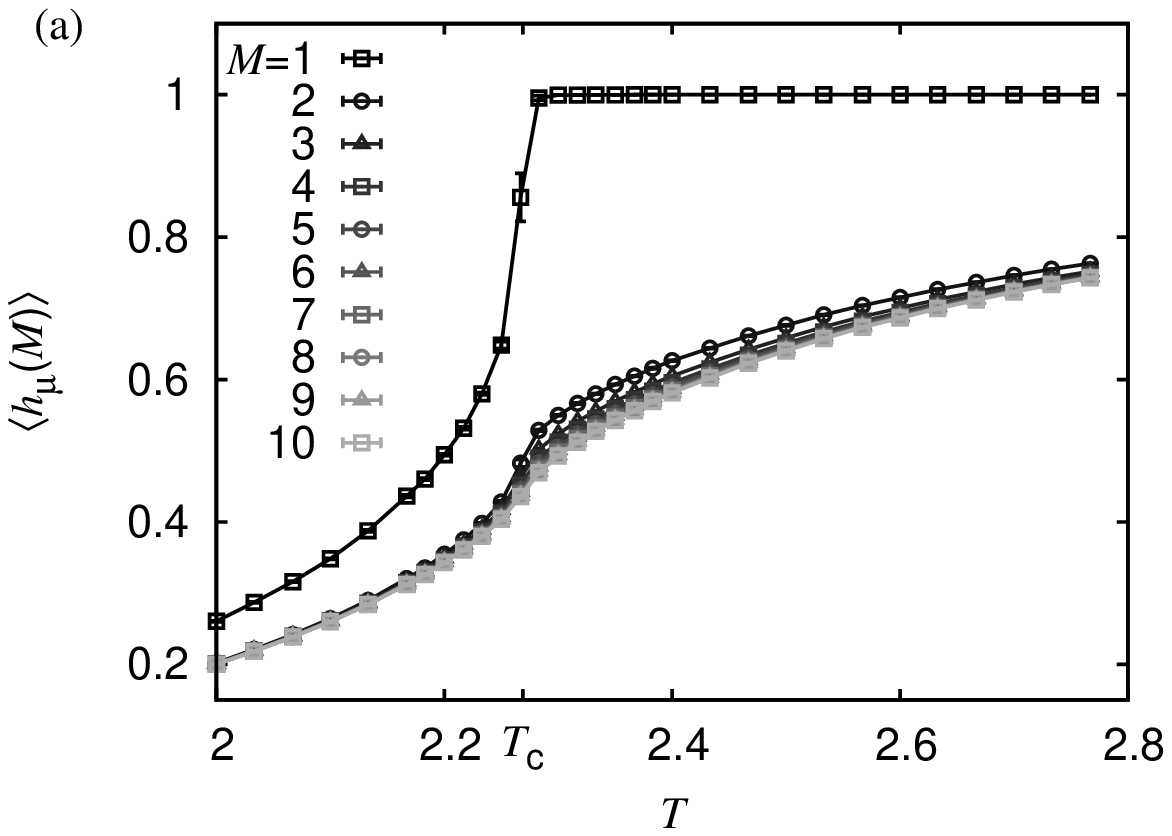}
\includegraphics[width=0.48\linewidth]{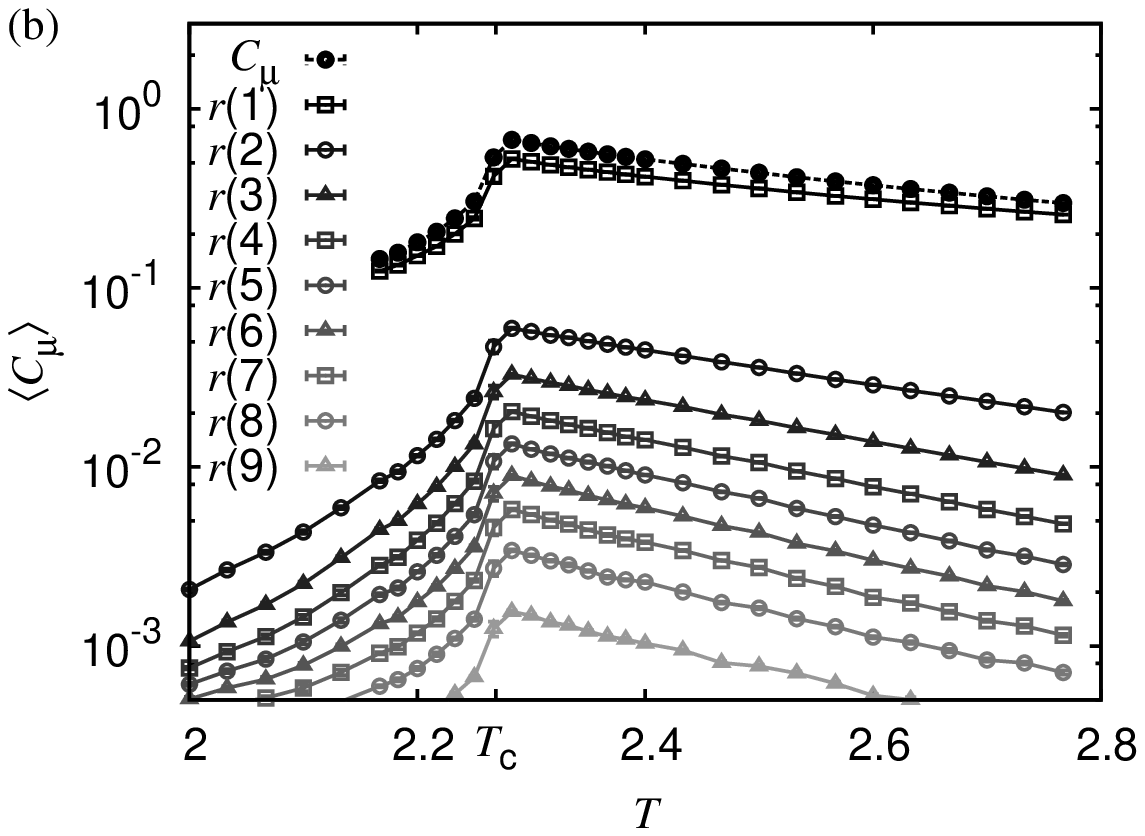}
\end{center}
\caption{
\label{fig:entropy_redundancy_convergence_temperature}
Results for the average entropy rate $h_\mu$ and the average complexity
$\langle C_\mu \rangle$ for binary sequences that describe the spin-flip dynamics
for the $2D$ Ising Ferromagnet at different temperatures $T$ considering sequences
of length $N=10^6$.
As explained in the text, the maximally feasible block-size for the computation of the
observables is set to $M=10$.
(a) convergence of the apparent entropy rates $\langle h_\mu(M) \rangle$ to the 
asymptotic entropy rate $\langle h_\mu\rangle$
For different block sizes $M=1,\ldots,10$. The results
for $M\ge 3$ fall almost on top of each other. Note that 
$\langle h_\mu \rangle$ only gives an upper bound on the actual entropy rate.
(b) average complexity $\langle C_\mu \rangle$ and contribution of the different per-symbol
$M$ redundancies $\langle r(M) \rangle$.
}
\end{figure}
%}}}2%%%%%%%%%%%%%%%%%%%%%%%%%%%%%%%%%%%%%%%%%%%%%%%%%%%
Finally, results for the numerical estimation of the entropy rate 
and the complexity considering $M_{\rm max}=10$ 
as a function of temperature $T$
are illustrated in Figs.\ \ref{fig:entropy_redundancy_convergence_temperature}(a) and (b), respectively.
The estimate of the average entropy rate $\langle h_\mu \rangle$, computed as explained above and shown in 
Fig.\ \ref{fig:entropy_redundancy_convergence_temperature}(a), will subsequently serve as a benchmark to 
which estimates that rely on the methods described in Sect.\ \ref{sect:patternDiscoveryAlgs} will be compared to.
The numerical derivative of the entropy rate with respect to the temperature indicates that 
right at the critical temperature, the increase of $\langle h_\mu \rangle$ is strongest (not shown). Further, the 
fluctuations $\chi_{h_\mu}\equiv \langle h_\mu^2 \rangle - \langle h_\mu \rangle^2$ exhibit an 
accentuated peak at $T_c$.
In the high-temperature paramagnetic phase, i.e.\ for all temperatures $T>T_c$, the single-symbol entropy rate assumes its extremal value $H(1)=\log_2(|\mathcal{A}|)=1$.
Also note that the complexity shown in Fig.\ \ref{fig:entropy_redundancy_convergence_temperature}(b) 
has an isolated peak close to the critical temperature. As evident from the latter figure, all terms in the sum 
of Eq.\ (\ref{eq:complexity})
 (i.e.\ the individual per-symbol $M$ redundancies) display a similar 
scaling behavior.

Again, note that part of these illustrating
results on the one-dimensional symbolic sequences 
reported above are qualitatively 
similar to those reported earlier in Ref.\ \cite{Arnold1996}. 
Conceptually similar analyses carried out on two-dimensional 
configurations of spins obtained from a simulation of the $2D$ Ising FM, 
reported in Ref.\ \cite{Feldman2008}, conclude that the excess entropy is
peaked at a temperature $T_c\approx 2.42$ in the paramagnetic phase slightly 
above the true critical temperature.
Similar results on the mutual information (which is equivalent to the excess entropy; 
see Ref.\ \cite{Crutchfield2003}) for the $2D$ Ising FM (and more general classical $2D$ spin models)
where recently presented in Ref.\ \cite{Wilms2011}. Therein, the authors conclude
that the mutual information reaches a maximum in the high-temperature paramagnetic
phase close to the system parameter $K=J/k_BT\approx0.41$ (for $J=k_B=1$ 
this corresponds to $T\approx 2.44$). Our new results and 
analyses,  which go
beyond the cited literature are presented in our main
result part Sec.\ \ref{sect:results}.

\subsection{An approximate measure of statistical complexity}
\label{ssect:approximateComplexity} 
Regarding the complexity, the convergence properties of the per-symbol $M$ redundancies as a function
of the block-size, displayed
in Fig.\ \ref{fig:entropy_redundancy_convergence}(b), suggest that $r(1)$ constitutes the dominant 
contribution to the sum in Eq.\ (\ref{eq:complexity}). As evident from 
the figure, $r(2)$ is approximately
one order of magnitude smaller than $r(1)$. In tandem with the observation that, pictured as a function
of temperature, $r(1)$ already has the shape characteristic for $C_\mu$ (see 
Fig.\ \ref{fig:entropy_redundancy_convergence_temperature}(b)) leads us to suggest $r(1)$ as 
an approximate estimator that might tell under which circumstances a given model exhibits a larger or smaller
complexity.
Using the fact that $h_\mu(1)\geq h_\mu$, we here define the \emph{approximate complexity} $c_\mu \in[0,1]$ as
\begin{eqnarray}
c_\mu \equiv r(1)/h_\mu(1) =  1-h_\mu/h_\mu(1). \label{eq:approxComplexity}
\end{eqnarray}
By definition, it is related to the complexity $C_\mu$ that quantifies the 
convergence properties of the entropy rate. 
Appealing to the definition of the per-symbol $M$ redundancies, $c_\mu$ quantifies the amount by which 
the entropy rate on the single-symbol level exceeds the asymptotic entropy rate.
To support intuition on the gross behavior of the approximate complexity note that the larger
the correlations between the symbols in a given sequence, the more patterns are missed by 
considering $h_\mu(1)$ in comparison to $h_\mu$, and the larger the numerical value of $c_\mu$ appears.
In the two limits of completely ordered and fully random symbol 
sequences, $c_\mu$ assumes a value of zero, respectively.
Note that Eq.\ (\ref{eq:approxComplexity}) is conceptually similar 
to the \emph{multi-information}
(given by the entropy rate difference between an elementary subsystem, i.e.\ a single spin, and
the infinite system)
introduced by Erb and Ay in Ref.\ \cite{Erb2004}. There, the authors considered the multi-information 
to characterize spatial spin-configuration for the $2D$ Ising FM in the thermodynamic limit by analytic means. 
Among other things, the authors conclude that the multi-information exhibits an isolated global maximum right 
at the critical temperature (see Theorem 3.3 of Ref.\ \cite{Erb2004}). 

The benefit of the approximate complexity is that if there is a way to numerically estimate the entropy 
rate $h_\mu$ (either as explained above, or by one of the algorithms introduced below in 
Sect.\ \ref{sect:patternDiscoveryAlgs}), there will immediately be a way to estimate also $h_\mu(1)$. 
The proposed measure of approximate complexity can even be computed by means of black-box data 
compression algorithms.
To facilitate intuition on that issue, consider a sequence $\myVec{S}$ of symbols that stems from 
the observation of a stochastic and ergodic process that possibly contains long-range correlations. 
Now, picture a black-box algorithm $A[\cdot]$ that upon postprocessing $\myVec{S}$
yields some estimate of the entropy rate, i.e.\ $h_\mu^{(A)} = A[\myVec{S}]$.
So as to pave the way towards an estimate of $h_\mu^{(A)}(1)$, consider the following:
for a process in which the values of the variables are independently and identically distributed,
i.e.\ for an \emph{IID process}, it holds that $h_\mu^{iid} = h_\mu^{iid}(1) = h_\mu^{iid}(2) = \ldots$.
For an IID process the block entropy rate grows linearly with the blocksize, and 
the associated complexity is zero (see Sect.\ V.A. on IID processes in Ref.\ \cite{Crutchfield2003}). 
An IID sequence related to the observed sequence $\myVec{S}$ is easily obtained as $\myVec{S}^{iid}=\pi[\myVec{S}]$, 
wherein $\pi[\cdot]$ signifies the permutation operator. Applying the permutation operator 
to the observed sequence destroys all patterns and yields an IID sequence with the same 
symbol frequencies as contained in $\myVec{S}$. In Ref.\ \cite{JimenezMontano2002b} this is referred 
to as ``standard random shuffle''. Consequently, an estimate of the single-symbol entropy rate 
using $A[\cdot]$ is provided by $h_\mu^{(A)}(1)=A[\pi[\myVec{S}]]$.
Note that for the Ising FM we find that at temperatures above the critical point 
it holds that $h_\mu(1)\approx 1$ (see Fig.\ \ref{fig:entropy_redundancy_convergence_temperature}(a)), 
hence we find $c_\mu=1-h_\mu$ for $T>T_c$.

%}}}1 %%%%

\section{Pattern discovery by means of symbolic substitution methods}
\label{sect:symbSubs}
%{{{1
\label{sect:patternDiscoveryAlgs}
As pointed out in the introduction, pair substitution methods like the NSRPS method, 
intending to quantify the degree of patterness of symbolic sequences, where introduced several decades ago 
by Ebeling and Jim\'enez-Monta\~no, see Ref.\ \cite{Ebeling1980}. 
The underlying elementary pair-substitution process is illustrated below in 
subsection 
\ref{ssect:NSRPS_intro}.
An algorithmic procedure that uses the NSRPS method in order to provide an elaborate
estimate of the entropy rate for a symbolic sequence is explained
in subsection \ref{ssect:NSRPS_entropyRate}.
Two further plans to approximately estimate the entropy rate are motivated in 
subsection 
\ref{ssect:NSRPS_algEntropyRate}.

\subsection{Non-sequential recursive pair substitution (NSRPS)}
\label{ssect:NSRPS_intro}

In order to describe an elementary non-sequential recursive pair substitution process,
consider a sequence $\myVec{S}^{(0)}=(s_1,\ldots,s_N)$, composed of $N$ symbols
stemming from a finite $m$-symbol alphabet $\mathcal{A}=\{a_i\}_{i=0}^{m-1}$.
Bear in mind that here, the initial sequences are assembled of symbols $s_i^{(0)}$ that 
stem from $\mathcal{A}=\{0,1\}$. 
The fundamental routine of the NSRPS algorithm, referred to as \emph{pair substitution},
might be illustrated as two step procedure:
\begin{enumerate}
\item[(i)] For a given sequence $\myVec{S}^{(0)}$, determine the frequency of all 
ordered pairs $e\in \mathcal{A}\times \mathcal{A}$ and 
identify $e_{\rm mfp}$, signifying the most frequent (ordered) length-two subsequence
of symbols. If the most frequent pair is not unique, signify one of them as $e_{\rm mfp}$. 
\item[(ii)] Construct a new sequence $\myVec{S}^{(1)}$ from $\myVec{S}^{(0)}$, wherein each 
full pattern $e_{\rm mfp}$ is replaced by a new symbol $a_m$. For this purpose, $\myVec{S}^{(0)}$ 
is scanned from left to right, and the existing alphabet $\mathcal{A}$ is augmented by $a_m$.
\end{enumerate}
This elementary pair substitution process might be executed iteratively. If one keeps track of the most 
frequent pairs of symbols and their substitutes at each iteration step, the initial sequence $\myVec{S}^{(0)}$
can be reconstructed any time.
This offers the possibility to design lossless compression algorithms based on
NSRPS.

%%%%%%%%%%%%%%%%%%%%%%%%%%%%%%%%%%%%%%%%%%%%%%%%%%%%%%%%
% FIGURE:       stopping criterion for NSRPS routine 
% LABEL :       fig:stopCrit_NSRPS
%%%%%%%%%%%%%%%%%%%%%%%%%%%%%%%%%%%%%%%%%%%%%%%%%%%%%%%%
%{{{2
\begin{figure}[t!]
\begin{center}
\includegraphics[width=0.48\linewidth]{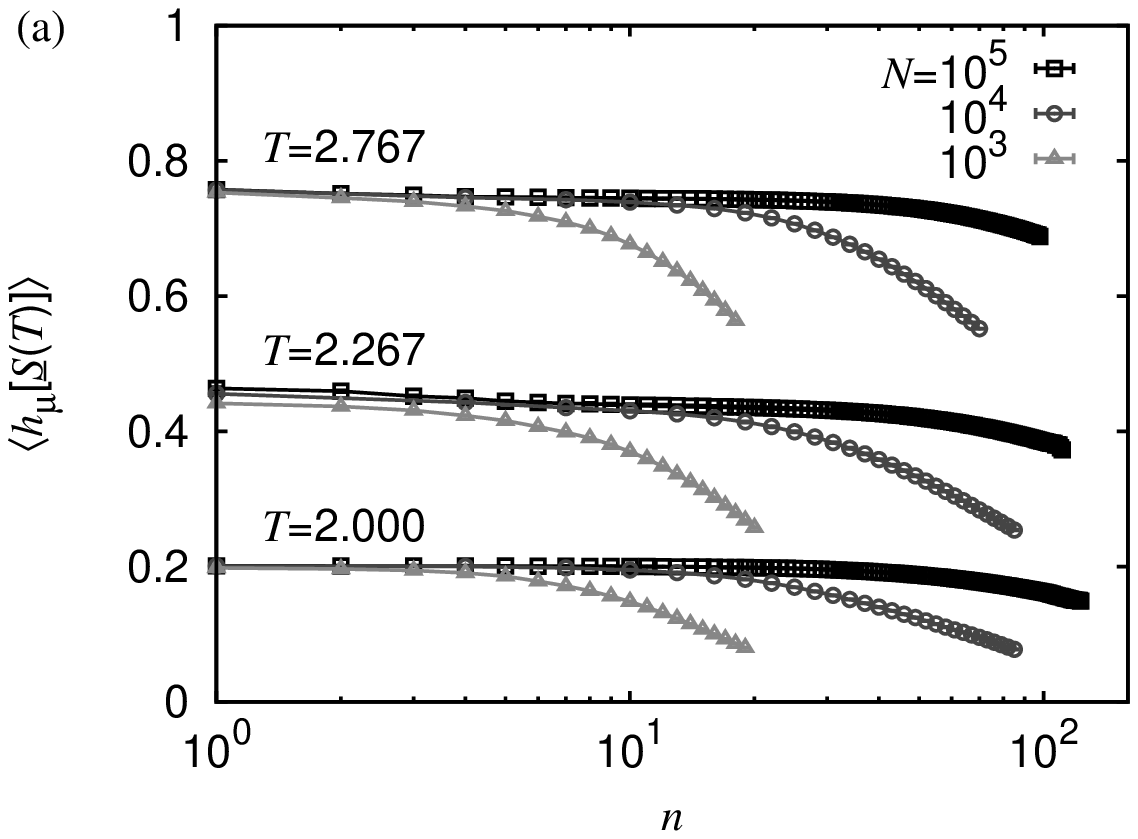}
\includegraphics[width=0.48\linewidth]{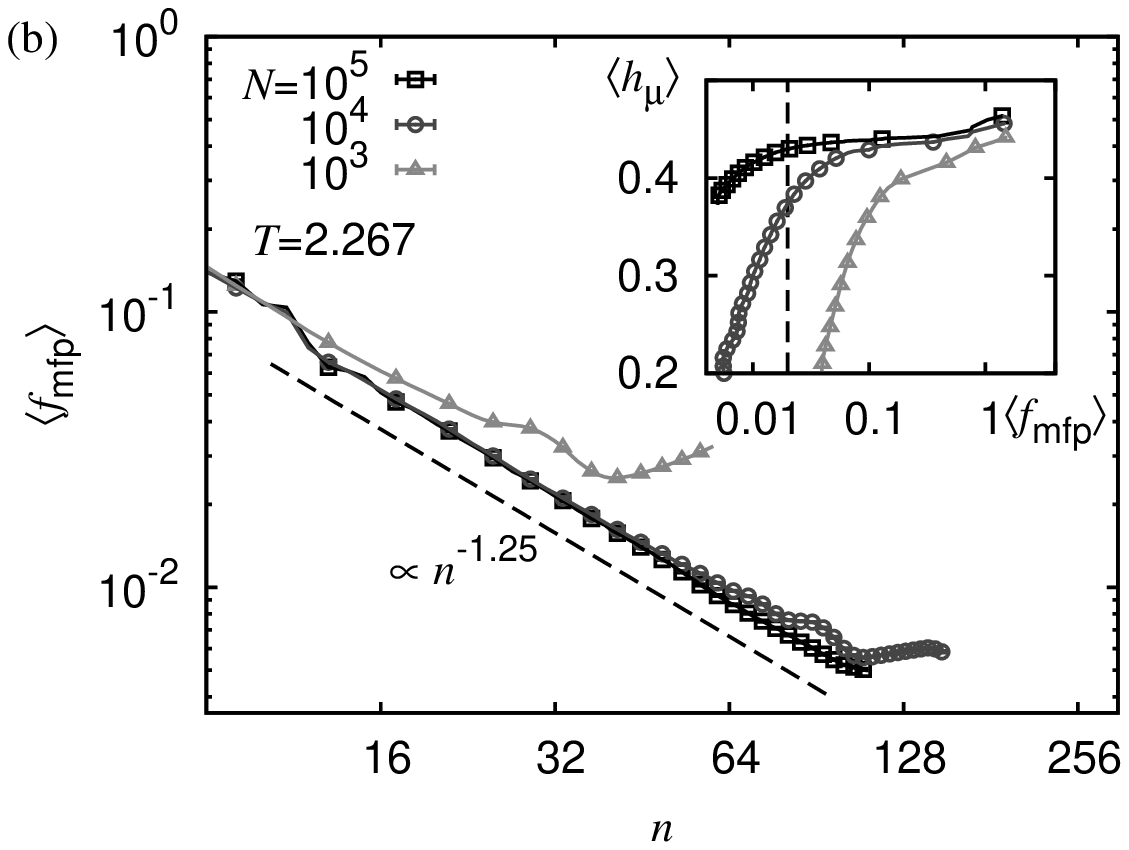}
\end{center}
\caption{
\label{fig:stopCrit_NSRPS}
Stopping criteria used to terminate the NSRPS procedure.
(a) Considering sequences at different temperatures $T$, the entropy
rate for sequence lengths $N>10^4$ converges to a plateau
at $\approx 5-20$ pair substitution steps. In order to 
compute the entropy rate via Eq.\ (\ref{eq:entropy_calcagnile}) 
we thus fix $N=10^5$ and consider a number of $n=20$ pair 
substitution steps.
(b) For sufficiently large sequence length the frequency 
$f_{\rm mfp}$ of the most frequent pair decreases with the number
of pairs substitution steps as $f_{\rm mfp}\propto n^{-1.25}$.
For other temperatures, the scaling behavior appears to be the same.
The inset illustrates the scaling behavior of the entropy rate 
as a function of the frequency $f_{\rm mfp}$. The dashed vertical 
line corresponds to the stopping condition $f_{\rm mfp}^{\rm min}$
used by Ref.\ \cite{Calcagnile2010}. 
}
\end{figure}
%}}}2%%%%%%%%%%%%%%%%%%%%%%%%%%%%%%%%%%%%%%%%%%%%%%%%%%%
\subsection{Numerical estimation of the entropy rate following Calcagnile \emph{et.\ al.}}
\label{ssect:NSRPS_entropyRate}
Based on the rigorous results reported in Ref.\ \cite{Benedetto2006}, and
further work by Ref.\ \cite{Calcagnile2010}, an elaborate entropy estimation 
algorithm based on the NSRPS method can be implemented.
Therein, the idea is that after a couple of pair substitutions, the most
frequent blocks, signifying the most common patterns up to a certain length 
within the sequence, are condensed into single symbols. Thus, it should by 
possible to find a good approximation to the actual entropy by considering the
$M$-block entropy (block entropy$=$BE) for small blocksize $M$ only. Following 
Ref.\ \cite{Calcagnile2010}, and denoting a symbolic sequence after a number
of $n$ pair substitution steps as $\myVec{S}^{(n)}$, a connection 
to the per-symbol entropy is available as
\begin{eqnarray}
h_\mu^{({\rm NSRPS-BE})}[\myVec{S}^{(0)}] = \lim_{n\to\infty}\lim_{N\to\infty} \frac{H(\myVec{S}^{(n)},2)-H(\myVec{S}^{(n)},1)}{\ell(\myVec{S}^{(0)})/\ell(\myVec{S}^{(n)}) },    \label{eq:entropy_calcagnile}
\end{eqnarray}
where $N$ refers to the length of the initial sequence $\myVec{S}^{(0)}$,
$H(\myVec{S},M)$ specifies the $M$-block entropy as estimated for the 
particular sequence $\myVec{S}$ and $\ell(\myVec{S})$ stands for the length 
of the sequence $\myVec{S}$ (note that $\ell(\myVec{S}^{(0)})=N$). 
From a point of view of numerical experiments, note that for a finite
value of $N$, the number $n$ of pair substitution steps is limited.
If the value for $n$ is chosen to be comparatively large and $\ell(\myVec{S}^{(n)})$
gets rather small, the statistics for the $M=2,1$-block entropies might become poor.
So as to cast Eq.\ (\ref{eq:entropy_calcagnile}) into a working algorithm 
for a  symbolic sequence of length $N$,
a proper stopping criterion for the iteration of the pair substitution step is required. 
Regarding that issue, the authors of Ref.\ \cite{Calcagnile2010} decided 
to stop the iteration of the pair substitution process as soon as the 
frequency of the most frequent pair gets smaller than $f_{\rm mfp}^{\rm min}=0.02$.
We here follow a slightly different approach that nevertheless yields quite
similar results. I.e., by considering sequences at different temperatures we monitored
the evolution of the entropy rate as function of the number of
pair substitution steps. Regardless of the temperature we 
found that for sufficiently long sequences ($N>10^4$) and after 
a number of approximately $5-30$ pair substitution steps, the entropy
rate converges to a plateau before it starts to decrease until
$h_\mu=0$ is reached. In Fig.\ \ref{fig:stopCrit_NSRPS}(a) this is illustrated for three
exemplary temperatures. Consequently, in order to assess the entropy
rate via Eq.\ (\ref{eq:entropy_calcagnile}) (following the approach of 
Calcagnile  \emph{et.\ al.}), we here fix the sequence length to $N=10^5$ and 
perform a number of $n=25$ pair substitution steps for all the
sequences considered.  
Regarding the frequency $f_{\rm mfp}$ of the most frequent pair we found that
for sequences not too short (i.e.\ sequence lengths $N > 10^4$),
it decreases algebraically with the number of pair substitution steps 
as $f_{\rm mfp}\propto n^{-1.25}$, see Fig.\ \ref{fig:stopCrit_NSRPS}(b). 
The stopping criterion of Ref.\ \cite{Calcagnile2010} would thus correspond 
to $n\approx 32$. The inset of Fig.\ \ref{fig:stopCrit_NSRPS}(b) shows that
for sequence lengths $>10^5$ the stopping condition $f_{\rm mfp}=0.02$ yields
an entropy value along a plateau immediately before the value of $h_\mu$ starts to decrease.
  
\subsection{Approximation of the entropy rate using further symbol substitution techniques}
\label{ssect:NSRPS_algEntropyRate}

The following subsection is based on the observation that data compression methods
allow to distinguish between regular and random sequences in the 
following sense:
A sequence that contains patterns (possibly on many scales) is not random 
but is compressible by means of symbol substitution methods.
Therein, a sequence $\myVec{S}$ is considered to be random if there exists no shorter 
sequence $\myVec{S}^\prime$ (written on the same alphabet as $\myVec{S}$)
that allows to construct $\myVec{S}$.
This further implies that the less patterns a sequence exhibits, the less compressible
it is. In terms of the sequences obtained for the $2D$ Ising Ferromagnet it is thus 
intuitive that a sequence recorded at $T\approx 0$ is highly compressible, whereas a
sequence recorded at $T\to\infty$ cannot be compressed much.
The sequences are called \emph{algorithmically simple} and \emph{algorithmically random}, respectively.

Now, consider an algorithm $M[\cdot]$ that returns some quantity describing how compressible
a given sequence appears to be. In this regard, the notion of algorithmically 
simple (random) shall translate to a small (large) value returned by $M[\cdot]$. 
Further, consider a length $N$ sequence recorded at finite temperature $T$ as well 
as a whole set of length $N$ sequences recorded at $T=\infty$.
A measure that might be used to approximate the entropy rate is given 
by the algorithmic entropy (AE), here defined as
\begin{eqnarray}
h_\mu^{({\rm AE})}[\myVec{S}(T)] = \frac{M[\myVec{S}(T)]}{\langle M[\myVec{S}(\infty)]\rangle}.  \label{eq:algEntropyRate}
\end{eqnarray}
The value of $\langle M[\myVec{S}(\infty)] \rangle$ is used to normalize the observable so 
that as $T\to0$ ($T\to\infty$) algorithmically simple (random) corresponds to $\langle h_\mu^{({\rm AE})}\rangle\to0$ 
($\langle h_\mu^{({\rm AE})}\rangle\to1$). Therein, the brackets $\langle\cdot\rangle$ denote
an average over different sequences.
Note that, by means of compression based estimators, it is possible to compute upper bounds
to the true per-symbol entropy, only. However, the aim of the presented subsection is not to provide
competitive estimators in comparison to those presented earlier in subsects.\ \ref{ssect:entropyRate_complexity_IsingFM} (BE) and \ref{ssect:NSRPS_entropyRate} (NSRPS-BE), but to prepare easy to compute
approximations to the approximate complexity (as reported later in subsects.\ \ref{ssect:NSRPS_results_entropyRate}
and \ref{ssect:results_systSize}).

\paragraph{Lempel-Ziv coding:}
If we consider data compression algorithms based on Lempel-Ziv coding, as, e.g., the
{\tt compress} algorithm contained in the {\tt zlib}-library \cite{zlib}, where $M[\cdot]$
returns the length of the compressed sequence, i.e.\ $M[\myVec{S}(T)]=\ell({\tt compress}[\myVec{S}(T)])$,
then $h_\mu^{({\rm AE})}$ effectively corresponds to the algorithmic entropy according to 
Lempel and Ziv as used by Ref.\ \cite{Ebeling1997} and we define 
\begin{eqnarray}
h_\mu^{({\rm ZLIB-AE})}[\myVec{S}(T)] = \frac{\ell({\tt compress}[\myVec{S}(T)])}{\langle\ell({\tt compress}[\myVec{S}(\infty)])\rangle}.  
\label{eq:algEntropyRate_zlib}
\end{eqnarray}
Note that Eq.\ (\ref{eq:algEntropyRate_zlib}) represents a fully 
data-compression based measure for the entropy rate.

\paragraph{NSRPS based symbol substitution:}
Recently, Nagaraj \emph{et.\ al.} detailed a method to measure the 
degree of randomness for symbolic sequences \cite{Nagaraj2011}. 
The idea behind their measure is as follows:
for a given sequence $\myVec{S}$ the pair substitution process might be iterated until the representation
of the sequence requires a single character, only. If this occurs after $N_{\rm ps}$
pair substitution steps, the corresponding sequence has zero entropy, i.e.\ 
$H(\myVec{S}^{(N_{\rm ps})},1)=0$, and $\myVec{S}^{(N_{\rm ps})}$ is called a \emph{constant} sequence.
In Ref.\ \cite{Nagaraj2011}, the minimal number of pair substitution steps $N_{\rm ps}$, needed to 
transform $\myVec{S}$ to a constant sequence, is adopted as
a measure of algorithmic randomness associated to the initial sequence.
As an example, consider the sequence $\myVec{S}=110010$. It consists of 
six symbols and exhibits the maximal single symbol entropy for a binary sequence.
A first application of the pair substitution routine identifies the most
frequent pair $e_{\rm mfp}=10$ and thus substitutes the respective 
subsequences by means of the new symbol $a_2=2$ (consequently the alphabet 
is amended to $\mathcal{A}=\{0,1,2\}$), to yield $\myVec{S}^{(1)}=1202$ having
$H(\myVec{S}^{(1)},1)=1.5$.
Finally, a repeated application of the pair substitution step on $\myVec{S}$ terminates 
after $N_{\rm ps}=4$ steps for the constant sequence $\myVec{S}^{(4)}=5$.
For the slightly modified sequence $\myVec{\tilde{S}}=101010$ (initially also having 
maximal entropy $H(\myVec{\tilde{S}},1)=1$),
the NSRPS algorithm terminates after just a single pair substitution process, where 
the constant sequence reads $\myVec{\tilde{S}}^{(1)}=222$.
One may now conclude that sequence $\myVec{S}$ exhibits a higher degree of randomness than 
the sequence $\myVec{\tilde{S}}$, since
the NSRPS algorithm requires a larger number of pair substitution steps in order 
to arrive at a constant sequence.
This is in accord with intuition, since, in contrast to the sequence $\myVec{S}=110010$, 
$\myVec{\tilde{S}}=101010$ exhibits a regular structure.
The value $N_{\rm ps}$ tells how well a given sequence might be compressed in terms of the NSRPS routine.
A small (large) value of $N_{\rm ps}$ indicates that the sequence is highly (hardly) compressible.
In accord with Eq.\ (\ref{eq:algEntropyRate}) we then define the 
NSRPS based algorithmic  entropy rate as 
\begin{eqnarray}
h_\mu^{({\rm NSRPS-AE})}[\myVec{S}(T)] = \frac{N_{\rm ps}[\myVec{S}(T)]}{\langle N_{\rm ps}[\myVec{S}(\infty)]\rangle}.  \label{eq:algEntropyRate_NSRPS}
\end{eqnarray}
Other than for the NSRPS-BE measure, explained in subsection \ref{ssect:NSRPS_entropyRate}, 
the NSRPS-AE measure requires no further tuning of a method-specific parameter.

%}}}1

\section{Results}
\label{sect:results}
%{{{1
Using the methods illustrated in the preceding section and for a range of 
temperatures including the critical point, we numerically 
compute the per-symbol entropies and approximate complexities for the $2D$ 
Ising Ferromagnet in subsection \ref{ssect:NSRPS_results_entropyRate}. 
In subsection \ref{ssect:results_systSize}, we then discuss the finite-size 
scaling behavior of the observables with respect to the system size $L$.
Further, we analyze the finite-size scaling behavior of the approximate complexity
in the sequence length $N$ for the data-compression based estimators 
NSRPS-AE and ZLIB-AE in subsection \ref{ssect:results_FSS_approximateComplexity}.
In subsection \ref{ssect:results_wolffClusterAlg} we report the 
results obtained for a spin-flip dynamics based on the Wolff cluster algorithm.
Finally, in subsection \ref{ssect:results_ceDiag}, we present the results obtained 
for the two different dynamics in purely information theoretic terms. 

\subsection{Numerical results for the entropy rate and approximate complexity}
\label{ssect:NSRPS_results_entropyRate}
%%%%%%%%%%%%%%%%%%%%%%%%%%%%%%%%%%%%%%%%%%%%%%%%%%%%%%%%
% FIGURE:       per-symbol entropy estimates obtained by the different methods
% LABEL :       fig:entropyEstimates_differentMethods
%%%%%%%%%%%%%%%%%%%%%%%%%%%%%%%%%%%%%%%%%%%%%%%%%%%%%%%%
%{{{2
\begin{figure}[t!]
\begin{center}
\includegraphics[width=0.48\linewidth]{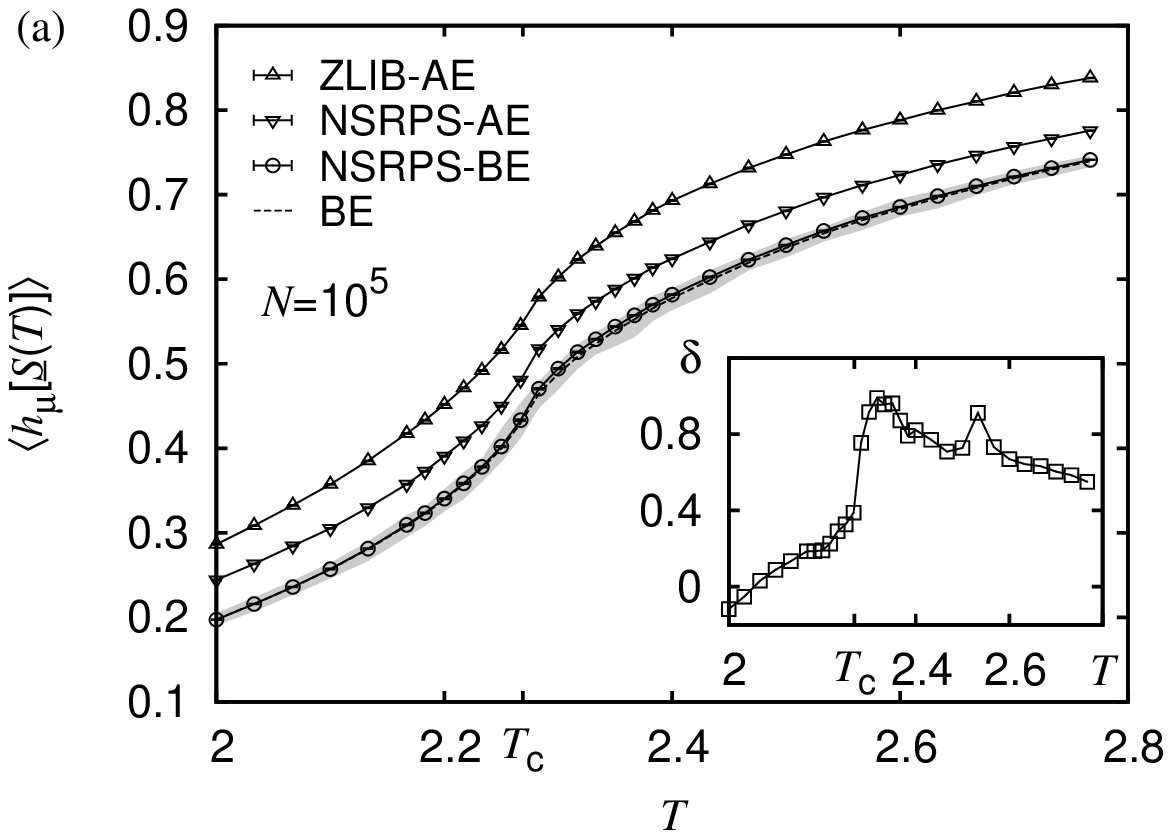}
\includegraphics[width=0.48\linewidth]{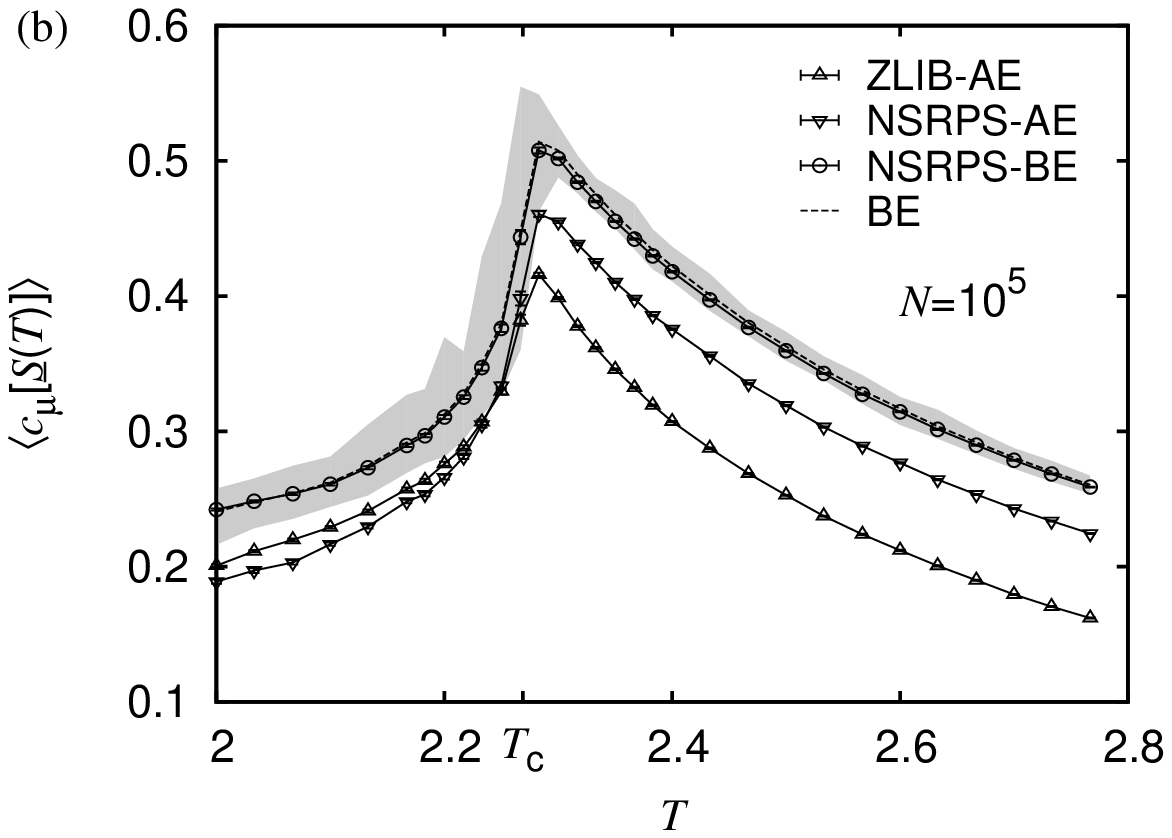}
\end{center}
\caption{
\label{fig:entropyComplexity_differentMethods}
Results for (a) the per-symbol entropy and (b) approximate complexity using different estimators for sequences
of length $N=10^5$. The results are averaged over a number of $100$ independent
sequences that characterize the spin-flip dynamics of the $2D$ Ising model simulated
via Metropolis dynamics at different temperatures $T$. 
The dashed line indicates the result for the average value 
obtained using the $M$-block entropy (BE) defined in Eq.\ (\ref{eq:mBlockSE}).
 The shaded are gives an account for
the difference between the maximal and minimal values of the entropy rate and approximate complexity
(in (a) and (b), respectively), obtained using the estimator BE.
As evident from the main plots, the NSRPS-BE estimator 
(see Eq.\ (\ref{eq:entropy_calcagnile}))
yields results in agreement with those of BE.
The compression based methods ZLIB-AE and NSRPS-AE (Eqs.\ 
(\ref{eq:algEntropyRate_zlib}) and
\ref{eq:algEntropyRate_NSRPS}, respectively) yield only upper (lower)
bounds to the true per-symbol entropy (approximate complexity). 
The inset in subfigure (a) shows the difference between the estimates
obtained via the NSRPS-BE and BE method in units of the standard error for the NSRPS-BE estimator (see text).
}
\end{figure}
%}}}2%%%%%%%%%%%%%%%%%%%%%%%%%%%%%%%%%%%%%%%%%%%%%%%%%%%

In Fig.\ \ref{fig:entropyComplexity_differentMethods}(a) we show the results for the per-symbol entropy 
$h_\mu$ obtained by using the different estimators introduced earlier.
As a benchmark we here consider the curve of $\langle h_\mu^{({\rm BE})}[\myVec{S}(T)]\rangle$, 
obtained for $100$ independent sequences of length $N=10^5$. For the computation of 
of an individual estimate $h_\mu^{({\rm BE})}[\myVec{S}]$ for a given sequences
$\myVec{S}$, the block size was restricted to $M\leq 10$. As pointed out
in Ref.\ \cite{JimenezMontano2002b}, upon analysis of an ensemble of 
independent finite sequences that stem from the same source, the per-symbol
entropies associated to the sequences are subject to statistical fluctuations.
So as to account for the spread of the per-symbol entropy among the $100$ independent
sequences at each value of $T$, the shaded area in the main plot indicates
the difference between the maximal and minimal value of the entropy rates thus obtained.
As evident from the figure, the entropy rates computed using the
NSRPS-BE method for $n=25$ compare quite well to 
the benchmark results. In this regard, the inset illustrates the difference
between the NSRPS-BE and BE measures in units of the standard deviation 
for the NSRPS-BE results, defined as 
\begin{eqnarray}
\delta(T) \equiv \frac{\langle h_\mu^{({\rm NSRPS-BE})}[\myVec{S}(T)] \rangle - \langle h_\mu^{({\rm BE})}[\myVec{S}(T)] \rangle}{{\rm sDev}(h_\mu^{({\rm NSRPS-BE})}[\myVec{S}(T)])}.
\label{eq:entropyDifference}
\end{eqnarray}
The inset shows that for $T>2.2$, the NSRPS-BE method systematically overestimates 
the benchmark curve. While the deviation increases for $T$ up to a value close to $T_c$,
it decreases as $T\to\infty$. Note that for the whole range of temperatures considered, the 
numerical estimates obtained using the NSRPS-BE and BE methods satisfy $|\delta(T)|<1$.
Further, an analysis for different values of $n$ reveals that as $n<20$ ($n>30$) and in the 
low (large) $T$ domain 
it holds that $|\delta(T)|>1$ (not shown).
The computation of the per-symbol entropy by means of the 
estimators $h_\mu^{({\rm ZLIB-AE})}$ and $h_\mu^{({\rm NSRPS-AE})}$ 
(defined in Eqs.\ (\ref{eq:algEntropyRate_zlib}) and
(\ref{eq:algEntropyRate_NSRPS}), respectively)
are less precise, see Fig.\ \ref{fig:entropyComplexity_differentMethods}(a), and will not be discussed further.

As explained earlier, the approximate complexity associated to a given sequence $\myVec{S}$
is computed as 
\begin{eqnarray}
c_\mu[\myVec{S}(T)] =1- \frac{h_\mu[\myVec{S}(T)]}{ h_\mu[\pi[\myVec{S}(T)]]}.  \label{eq:approxCplx_sequence}
\end{eqnarray}
The numerical results, obtained for the approximate complexities considering $100$
independent sequences of length $N=10^5$ are presented in Fig.\ \ref{fig:entropyComplexity_differentMethods}(b).
Again, we consider the averages obtained by means of the BE method as benchmark to which 
the other methods are compared to. As for the entropy rated considered above, the results obtained 
using the NSRPS-BE method yields an approximate complexity which compares well to the
BE estimate. However, note that for temperatures $T\gtrapprox 2.1$ the NSRPS-BE estimate
overestimates the BE estimate systematically. 
The results obtained by means of the 
estimators $c_\mu^{({\rm ZLIB-AE})}$ and $c_\mu^{({\rm NSRPS-AE})}$ 
(computed similar to Eq.\ (\ref{eq:approxCplx_sequence}))
only yield a crude lower bound on the approximate complexity, as compared to the more sophisticated 
estimates.

As defined here, the numerical value of $c_\mu$ lies in between 
zero and one and is small
at low and high temperatures, where the sequences are 
algorithmically simple and random, respectively.
At intermediate temperatures, i.e.\ in the vicinity of the critical 
temperature, long range correlations between symbols in the sequences 
appear, resulting in a comparatively large value of $c_\mu$.
Thus, the quantity $c_\mu$ behaves as we expect it for
a quantity indicating ``complexity'' of a sequence.
Note that albeit the magnitude of the approximate complexity at a given temperature for the 
various estimator might differ, the peak positions
of the different curves are all locate close by the critical 
temperature $T_c\approx 2.269$.

%%%%%%%%%%%%%%%%%%%%%%%%%%%%%%%%%%%%%%%%%%%%%%%%%%%%%%%%
% FIGURE:       NSRPS-BE, NSRPS-AE, ZLIB-AE: finite size scaling
% LABEL :       fig:results_diffL
%%%%%%%%%%%%%%%%%%%%%%%%%%%%%%%%%%%%%%%%%%%%%%%%%%%%%%%%
%{{{2
\begin{figure}[t!]
\begin{center}
\includegraphics[width=0.48\linewidth]{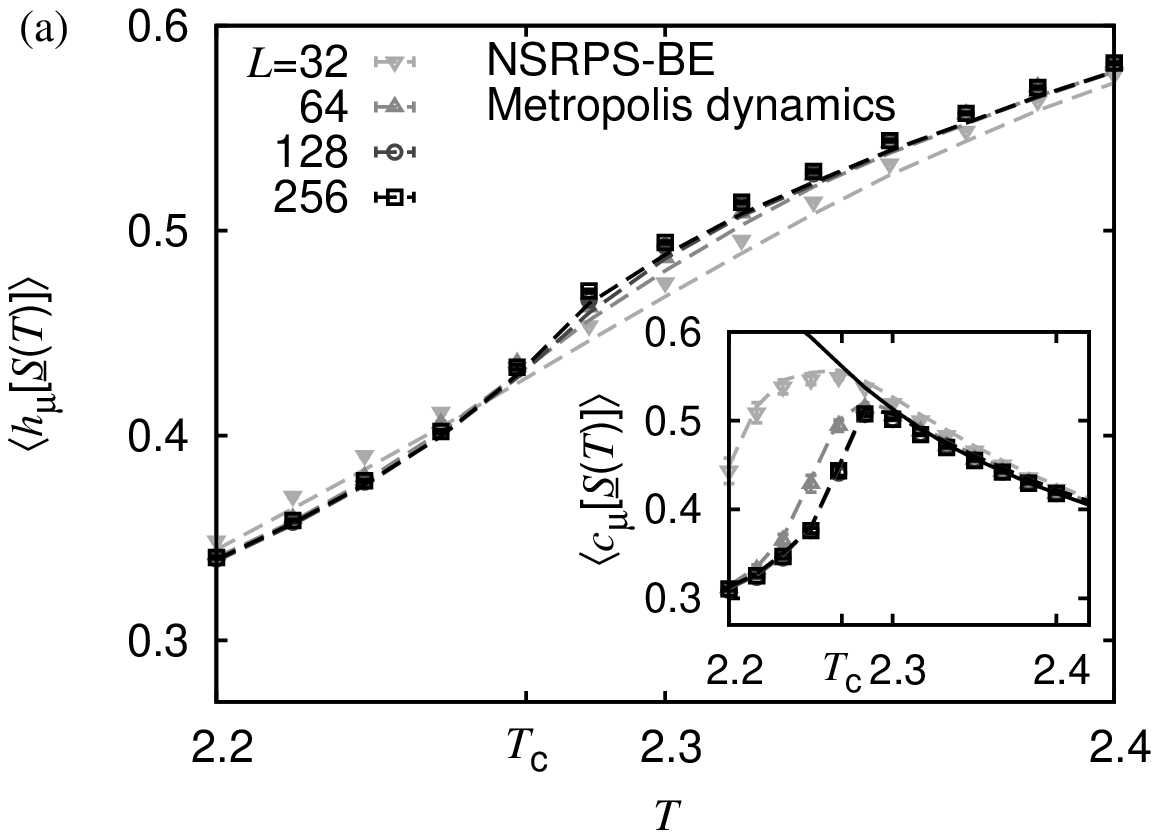}
\includegraphics[width=0.48\linewidth]{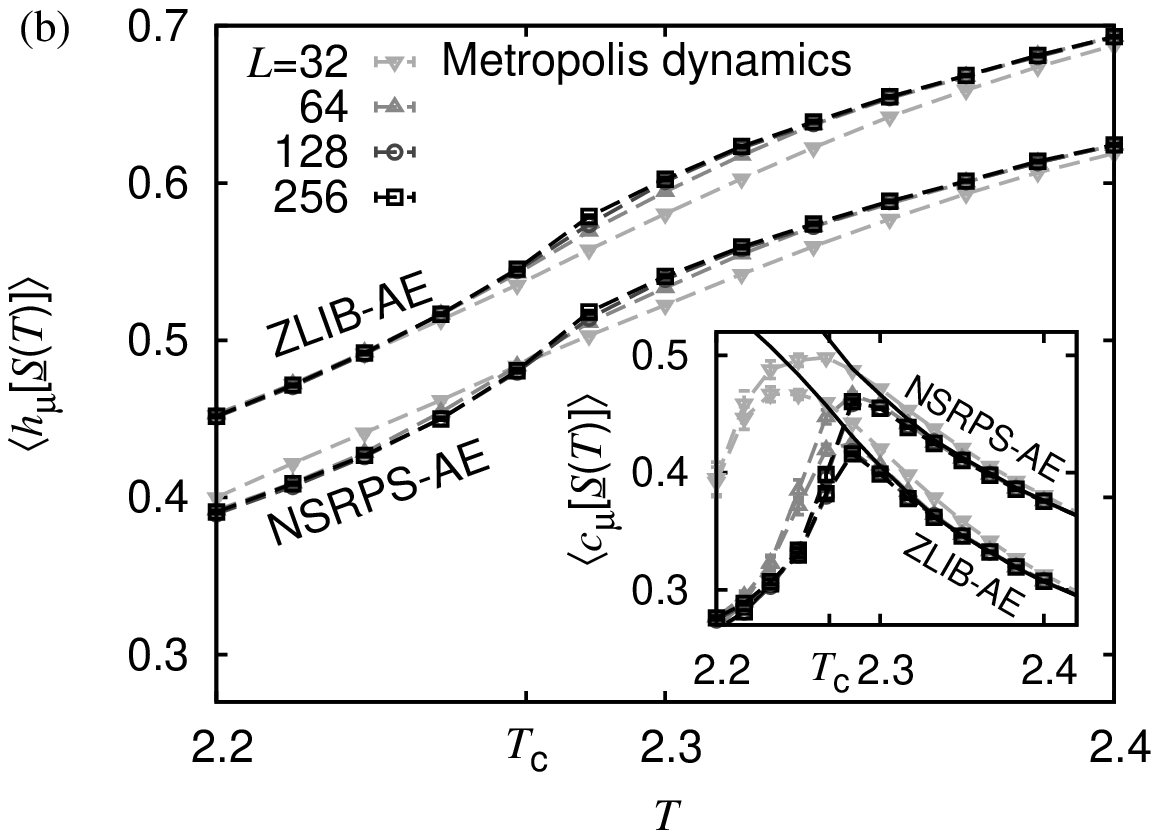}
\end{center}
\caption{
\label{fig:results_diffL}
System size dependence of the per-symbol entropy (main plot) and 
approximate complexity (inset). The sequence length is fixed to $N=10^5$.
(a) shows the results obtained using the NSRPS-BE method (data points) and 
BE method (dashed lines) introduced
in subsects.\ \ref{ssect:NSRPS_entropyRate} and \ref{ssect:notation}, respectively.
(b) shows the results obtained using the NSRPS-AE and ZLIB-AE methods
introduced in subsection \ref{ssect:NSRPS_algEntropyRate} (here, the dashed lines are 
just guides to the eye). 
In either case, the solid line in the inset (without symbols)
 illustrates the curve $1-\langle h_\mu[\myVec{S}(T)]\rangle$ for $L=64$.
}
\end{figure}
%}}}2%%%%%%%%%%%%%%%%%%%%%%%%%%%%%%%%%%%%%%%%%%%%%%%%%%%
\subsection{Finite-size scaling regarding the system size}
\label{ssect:results_systSize}

An analysis of the per-symbol entropy for sequences of length $N=10^5$, obtained for 
systems of different sizes $L=32$ through $256$ and for the different estimation methods,
is shown in the main plots of Figs.\ \ref{fig:results_diffL}(a) and (b).
Regarding the elaborate NSRPS-BE method, 
the curves for different system sizes have a common crossing 
point close to $T_c$, see Fig.\ \ref{fig:results_diffL}(a). At a given temperature below (above) the critical point, 
increasing $L$ results in a smaller (larger) value of the per-symbol entropy.
At $T\to0$ as well as for $T\to\infty$, the data points for the different system sizes
coincide (note that the figure only shows a zoom-in on the interval $T\in[2.2,2.4]$, enclosing $T_c$). 
While it is possible to distinguish the data curves for $L=32$ and $L=256$, it is hard
to tell apart the curves for $L=128$ and $L=256$. This might indicate that the 
data curves at $L=128$ and $256$ are reasonable approximations to the thermodynamic limit $L\to\infty$.
For the approximate complexity shown in the inset of Fig.\ \ref{fig:results_diffL}(a) 
we find that its peak gets more pronounced as $L$ increases.
Further, for the smallest system size, i.e.\ $L=32$, the peak is located slightly
below $T_c$, shifting towards a higher temperature as $L$ increases.
Again, it is hard to tell apart the data curves for $L=128$ and $256$. 
The dashed line in the inset illustrates the curve
$1-\langle h_\mu[\myVec{S}(T)]\rangle$ for $L=64$, which compares well 
to the scaling of $\langle c_\mu[\myVec{S}(T)]\rangle$ as $T>T_c$
(as pointed out in subsection \ref{ssect:approximateComplexity} this is due to the fact that 
$\langle h_\mu[\pi[\myVec{S}(T)]]\rangle\approx 1$ for temperatures $T>T_c$).
While similar observations can be made for the NSRPS-AE method, the finite-size scaling for the ZLIB-AE method 
is different. As can be seen from Fig.\ \ref{fig:results_diffL}(b), data curves obtained using the
ZLIB-AE method exhibit no crossing point at all. However, still there is the tendency that 
for temperatures above $T_c$, an increasing system size leads to an increasing value of the 
per-symbol entropy.

%%%%%%%%%%%%%%%%%%%%%%%%%%%%%%%%%%%%%%%%%%%%%%%%%%%%%%%%
% FIGURE:       peak scaling, complexity - NSRPS-AE 
% LABEL :       fig:peakScaling_complexity_nsrpsAE
%%%%%%%%%%%%%%%%%%%%%%%%%%%%%%%%%%%%%%%%%%%%%%%%%%%%%%%%
%{{{2
\begin{figure}[t!]
\begin{center}
\includegraphics[width=0.48\linewidth]{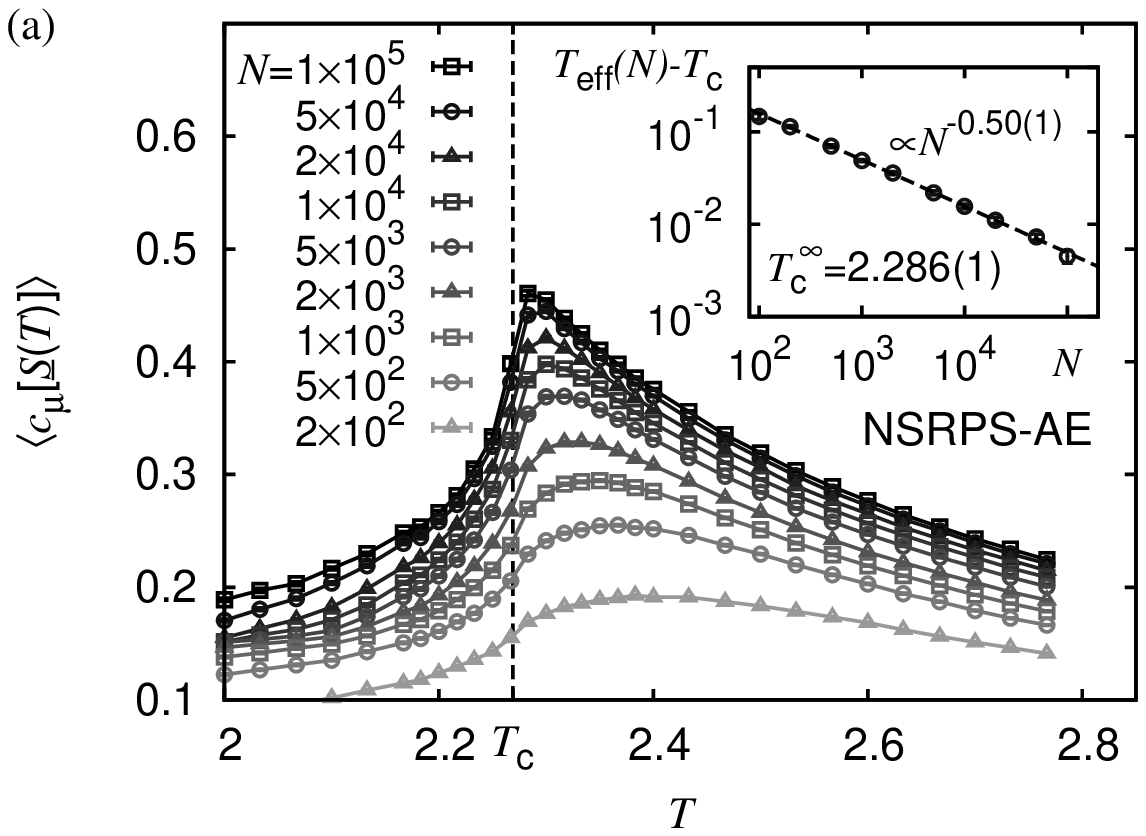}
\includegraphics[width=0.48\linewidth]{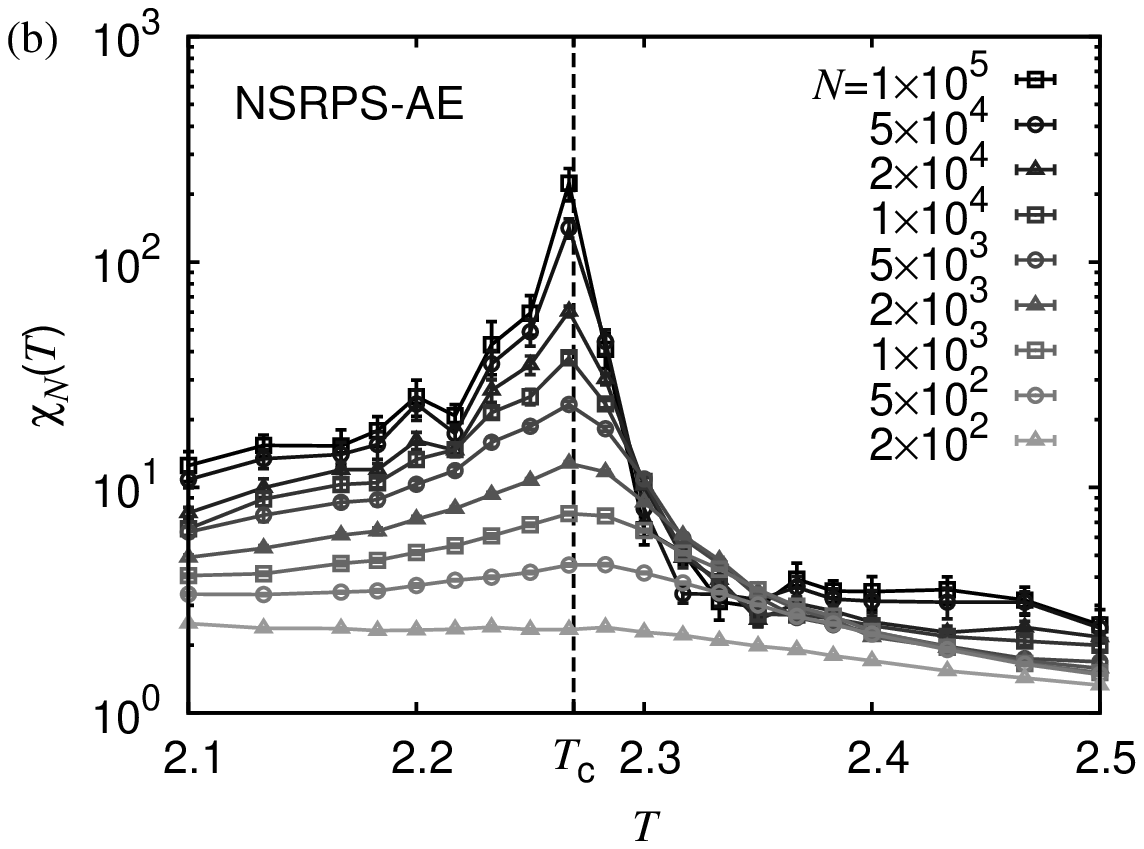}\\
\includegraphics[width=0.48\linewidth]{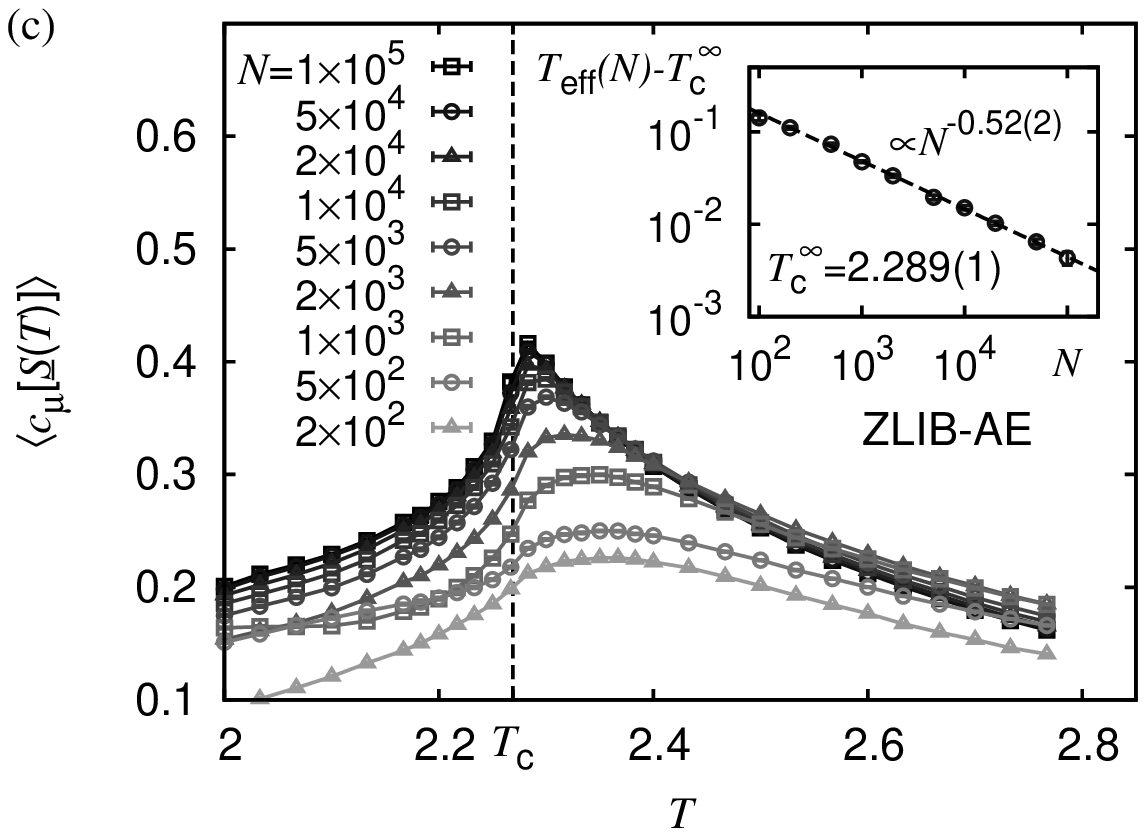}
\includegraphics[width=0.48\linewidth]{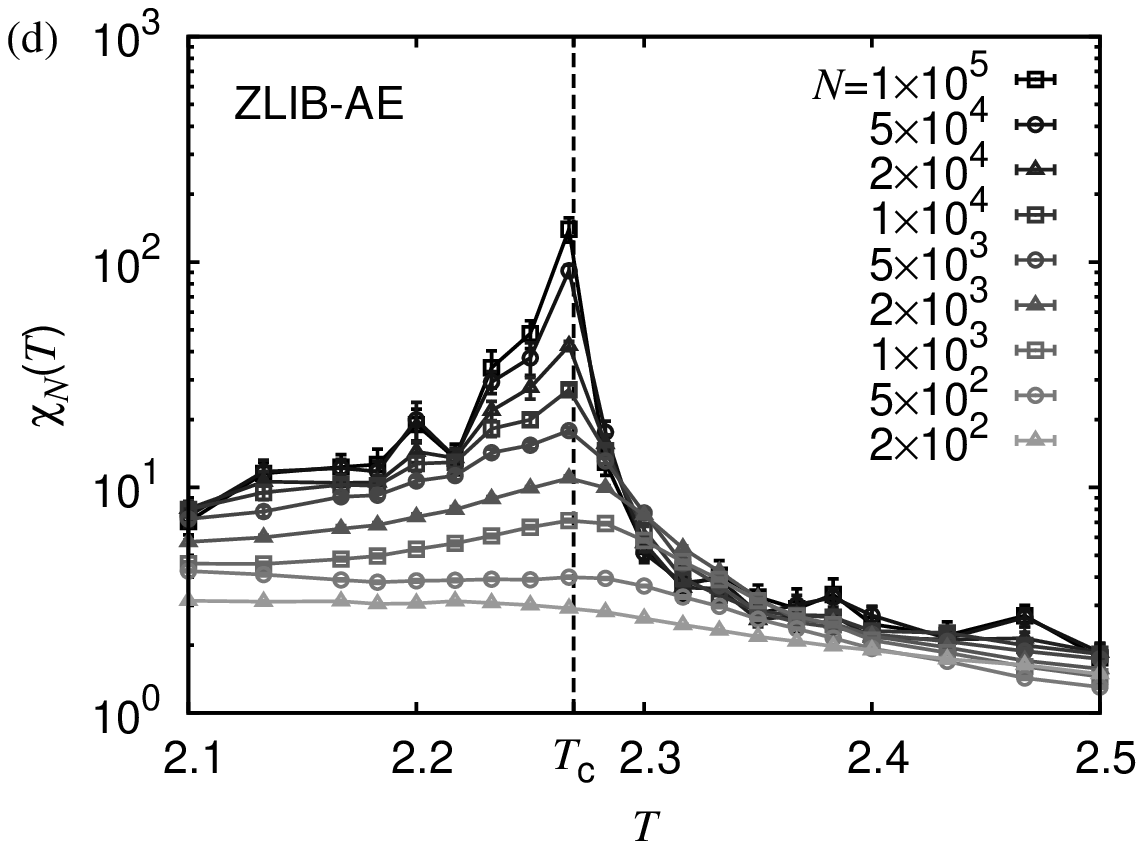}
\end{center}
\caption{
\label{fig:peakScaling_complexity_nsrpsAE}
Results of the finite-size scaling analysis for the peak-location considering
the approximate complexity for different sequence lengths $N$
at fixed system size $L=256$.
(a) illustrates the data curves obtained by means of the NSRPS-AE method and
the inset shows the scaling of the associated peak locations.
(b) indicates the scaling behavior of the finite-size fluctuations for 
the approximate complexities.
The subfigures (c) and (d) show the same data as (a) and (b), obtained 
using the ZLIB-AE method. Lines are guides to the eyes only.
}
\end{figure}
%}}}2%%%%%%%%%%%%%%%%%%%%%%%%%%%%%%%%%%%%%%%%%%%%%%%%%%%
\subsection{Sequence length dependence of the approximate complexity}
\label{ssect:results_FSS_approximateComplexity}

A relevant parameter that controls how well the statistics of the spin-flip dynamics is captured
by the analyzed sequences is the sequence length $N$. The longer the sequence, the more
patterns might be resolved and the better the approximation of the statistical properties
of the spin-flip dynamics.  
In this regard, we performed a finite-size scaling analysis for the peak-location of the
approximate complexity as obtained by the NSRPS-AE and ZLIB-AE measures.
We did not perform such an analysis for the NSRPS-BE method, since, as evident from Fig.\ 
\ref{fig:stopCrit_NSRPS}, the convergence properties of the entropy rate render it hard to yield reliable
results for sequence lengths $N<10^4$. 

Considering the NSRPS-AE estimator, Fig.\ \ref{fig:peakScaling_complexity_nsrpsAE}(a) 
indicates that for increasing values of $N$ 
the peak-position approaches the critical value $T_c$ from above.
The same holds also for the ZLIB-AE estimator, see Fig.\ \ref{fig:peakScaling_complexity_nsrpsAE}(c).
As evident from the figures, the curves for the different sequence lengths 
$N$ are peaked at effective critical points 
$T_c^{\rm eff}(N) > T_c$. The peaks get more pronounced as $N$ increases.
By fitting 5th-order polynomials to the data curves in order to
estimate the precise location of the peaks (where errorbars are obtained
via bootstrap resampling \cite{practicalGuide2009}), we found that 
the effective critical points exhibit the scaling behavior 
$T_c^{\rm eff}(N) = T_c^{\infty} + a\cdot N^{-b}$, see inset of Figs.\ 
\ref{fig:peakScaling_complexity_nsrpsAE}(a),(c).
This corresponds to standard finite-size scaling near phase transitions
\cite{goldenfeld1992} 
when changing the system size $L$, which is instead kept fixed here.
The fit parameters obtained for the NSRPS-AE (ZLIB-AE) method read 
$T_c^{\infty}=2.286(1)$ and $b=0.50(1)$ ($T_c^{\infty}=2.289(1)$ and 
$b=0.52(2)$). Note that this exponent is different from the
scaling observed when varying the system size, where the exponent $-1/\nu=-1$ 
is relevant and related to the correlation length exponent $\nu=1$.
For increasing sequence length, the approximation of the peak-position 
by means of the 5-th order polynomials gets rather imprecise, which might
account for the deviation between $T_c^{\infty}$ and the true critical temperature
$T_c$.
Accordingly, we performed a further analysis considering the NSRPS-AE method for 
sequences up to length $N=5000$ only (where the peaks can be fit well), resulting 
in the improved estimate $T_c^{\infty}=2.270(5)$.
Note that in any case, the value of $T_c^{\infty}$ is reasonably close to the 
critical temperature, and that these results are obtained at fixed $L=256$ by 
varying the length $N$ of the sequences. Further, the finite-size fluctuations 
$\chi_N(T)=N \times {\rm var}(c_\mu[\myVec{S}(T)])$, shown in 
Figs.\ \ref{fig:peakScaling_complexity_nsrpsAE}(b),(d) for NSRPS-AE and ZLIB-AE, respectively, exhibit a peak 
that also tends towards $T_c$ as $N$ increases. Thereby, once the sequence length
exceeds $N\approx 5\times 10^3$, the peaks are located right at 
the critical temperature. These results show that the methods
we used to estimate the ``complexity'' of a sequence are not 
only giving qualitatively
satisfying results but can be used for rather precise quantitative
estimates from finite-size system data.

\subsection{Spin-flip dynamics based on the Wolff cluster algorithm}
\label{ssect:results_wolffClusterAlg}

%%%%%%%%%%%%%%%%%%%%%%%%%%%%%%%%%%%%%%%%%%%%%%%%%%%%%%%%
% FIGURE:       NSRPS-BE: system size scaling WOLFF 
% LABEL :       fig:results_L_Wolff
%%%%%%%%%%%%%%%%%%%%%%%%%%%%%%%%%%%%%%%%%%%%%%%%%%%%%%%%
%{{{2
\begin{figure}[t!]
\begin{center}
\includegraphics[width=0.6\linewidth]{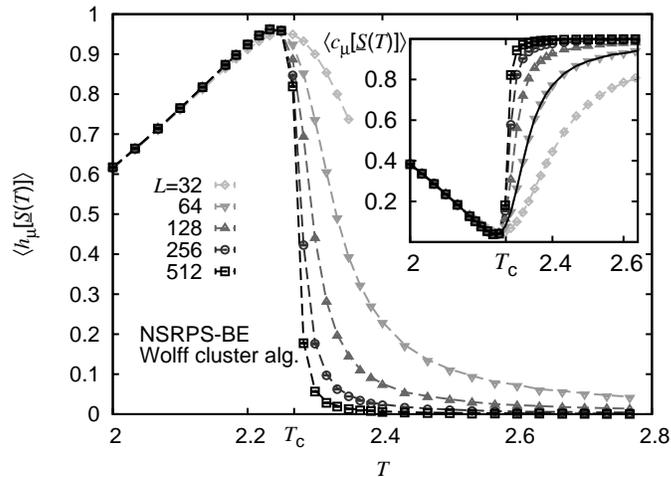}
\end{center}
\caption{
\label{fig:results_L_Wolff}
Results of the finite-size scaling analysis for the per-symbol entropy (main plot) and 
complexity (inset) considering different 
system sizes $L$ for a spin-flip dynamics based on the Wolff cluster algorithm. 
The solid line in the inset illustrates
the curve $1-\langle h_\mu[\myVec{S}(T)]\rangle$ for $L=64$.
In the figure, data points correspond to the estimate obtained using the NSRPS-BE method,
while the dashed lines indicate the respective results obtained using the BE method. 
}
\end{figure}
%}}}2%%%%%%%%%%%%%%%%%%%%%%%%%%%%%%%%%%%%%%%%%%%%%%%%%%%

In Fig.\ \ref{fig:results_L_Wolff} we show the results for the entropy rate and
approximate complexity, where, instead of a single spin-flip MC simulation using
Metropolis-dynamics (as above) we used the Wolff-cluster-algorithm \cite{wolff1989}.
The time-unit within these simulation is given by a single cluster-construction process
(in contrast: using the single spin-flip MC simulation for a square lattice of side length $L$, 
the time unit consists of a number of $L\times L$ independent spin-flip attempts for randomly
chosen spins, comprising one sweep). 
The Wolff-cluster-algorithm is most efficient at $T_c$, yielding 
binary sequences where subsequent symbols are effectively uncorrelated,
as reflected by the minimum of the approximate complexity, where 
$\langle c_\mu[\myVec{S}(T_c)]\rangle$ assumes a small value close to zero.
Further, the fact that $\langle h_\mu[\myVec{S}(T)]\rangle$ exhibits a peak value 
of nearly one at $T_c$ indicates that, at the critical temperature, 
the sequences appear to be algorithmically random.
Hence, although the system being simulated is the same,
the strong differences in the resulting entropy and complexity curves
are easily understood by the dynamics of the algorithm.

The system size dependence of the per-symbol entropy, illustrated in
the main plot of Fig.\ \ref{fig:results_L_Wolff}, can be understood 
in terms of the cluster-construction characteristics of the Wolff cluster 
algorithm.
At a temperature $T<T_c$, a cluster constructed at a given time-step comprises
almost all spins on the lattice, regardless of the system size (in the limit $T\to0$, 
a cluster comprises all spins). It is thus very 
likely that a given spin is contained in that cluster and gets flipped frequently.
At temperatures $T>T_c$, the clusters have some typical size that does not depend
on the size of the system (in the limit $T\to\infty$, a cluster consists of a single 
spin only). Considering a fixed temperature $>T_c$, the larger the system 
size $L$, the smaller the relative size of a (typical) cluster appears. 
Consequently, for a given spin it is less likely to be contained in a cluster 
as $L$ increases and the spin gets flipped only rarely. Thus, the flip-frequency
decreases upon increasing system size.

Further, at temperatures below $T_c$ where a given spin flips rather 
frequently, we find $\langle h_\mu[\pi[\myVec{S}(T)]]\rangle \approx 1$. 
Hence, for $T<T_c$ it holds rather precisely that 
$\langle c_\mu[\myVec{S}(T)]\rangle \approx 1-\langle h_\mu[\myVec{S}(T)]\rangle$.
For $T>T_c$, $\langle h_\mu[\pi[\myVec{S}(T)]]\rangle$ decreases 
slightly with increasing temperature (not shown). The decrease is monotonous and the observable 
takes values in between $1$ and $0.8$, so we still find that the above relation
holds approximately.
This is also evident by visually inspecting the data curves for the per-symbol entropy and
approximate complexity displayed in Fig.\ \ref{fig:results_L_Wolff}.
In the inset, the dashed line illustrates $1-\langle h_\mu[\myVec{S}(T)]\rangle$ for $L=64$,
which compares well to the respective data curve showing $\langle c_\mu[\myVec{S}(T)]\rangle$. 
This leads us to suggest that for the spin-flip dynamics induced by the Wolff cluster algorithm, 
$h_\mu$ and $c_\mu$ are trivially correlated.

In general, one should find that as $T\to\infty$ the results for the dynamics using 
the Wolff cluster algorithm should match those of the single spin-flip Metropolis dynamics.
However, as a comparison of Figs.\ \ref{fig:results_diffL} and \ref{fig:results_L_Wolff} indicates, the 
results for the per-symbol entropies and approximate complexities in the limit $T\to\infty$ 
for the different dynamics are completely different.  
This difference is solely due to the definition of a time-unit regarding the two spin-flip dynamics.
By means of a proper rescaling of the time-unit related to the Wolff-cluster-algorithm 
we verified that the sequences supplied by both spin-flip dynamics yield similar results as $T\to \infty$.
In particular at $T=10.0$ and for a comparatively small sequence length of $N=10^3$ 
we obtained $\langle  c_\mu[\myVec{S}(T)] \rangle=0.022(2)$
and $\langle c_\mu[\myVec{S}(T)] \rangle=0.023(7)$ for Metropolis dynamics and
Wolff-cluster algorithm using the NSRPS-AE method, respectively (we further checked
that for sequence-lengths $N\geq 10^4$ the NSRPS-BE method yields approximately 
the same results).

\subsection{Aproximate Complexity-Entropy diagrams for the different spin-flip dynamics}
\label{ssect:results_ceDiag}

%%%%%%%%%%%%%%%%%%%%%%%%%%%%%%%%%%%%%%%%%%%%%%%%%%%%%%%%
% FIGURE:       approx. complexity-entropy diagram for different dynamics
% LABEL :       fig:results_ceDiag
%%%%%%%%%%%%%%%%%%%%%%%%%%%%%%%%%%%%%%%%%%%%%%%%%%%%%%%%
%{{{2
\begin{figure}[t!]
\begin{center}
\includegraphics[width=0.6\linewidth]{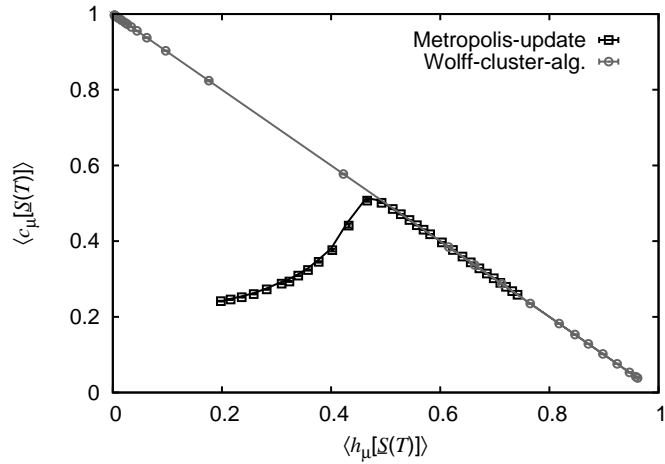}
\end{center}
\caption{
\label{fig:results_ceDiag}
Approximate complexity-entropy diagram obtained 
for a spin-flip dynamics based on single-spin Metropolis update and the Wolff cluster algorithm. 
In the analysis, we considered symbolic sequences $\myVec{S}$ of length $N=10^5$ at 
different temperatures $T$ (the relation between temperature and entropy is illustrated
in Fig.\ \ref{fig:entropyComplexity_differentMethods}(a)).
In the figure, data points correspond to the results obtained using the NSRPS-BE method,
while the solid lines indicate the respective results obtained using the BE method. 
The figure characterizes the different dynamics in purely information-theoretic coordinates.
}
\end{figure}
%}}}2%%%%%%%%%%%%%%%%%%%%%%%%%%%%%%%%%%%%%%%%%%%%%%%%%%%

In the subsections above, we illustrated the information-theoretic observables ``entropy rate'' and
``approximate complexity'' as function of the model parameter $T$. In order to account 
for a completely information-theoretic characterization of the different spin-flip dynamics, 
we illustrate the corresponding (approximate) complexity-entropy diagrams in Fig.\ \ref{fig:results_ceDiag}. 
As evident from the figure, the relation between the information-theoretic coordinates 
depends on the underlying spin-flip dynamics.
In case of the single-spin flip Metropolis update the dynamics close to the critical temperature 
suffers from severe slowing down due to long-range correlations.
Off criticality, correlations are less strong.
Consequently, the approximate complexity (which effectively accounts for correlations) is peaked at an
entropy value $\approx 0.5$, reflecting the critical temperature. 
In terms of this very simple update mechanism, the evolution of the $2D$ Ising FM appears to be 
highly intricate.
On the contrary, the elaborate dynamics provided by the Wolff cluster algorithm makes the 
evolution of the $2D$ Ising FM at the critical temperature maximally efficient. 
It does not suffer from critical slowing down and successive spin configurations are 
effectively uncorrelated. 
Consequently, the time series related to the orientation of a particular spin on the lattice
appears to be maximally random at $T_c$.
Hence, the critical temperature is reflected by a large entropy rate and low approximate 
complexity. 
The region of very low entropy rate and very large complexity indicates periodic sequences,
obtained at small temperatures $T$ where at each time step almost all spins are flipped. 
Further, as pointed out in subsection \ref{ssect:results_wolffClusterAlg}, Fig.\ \ref{fig:results_ceDiag} 
shows that for the dynamics provided by the Wolff cluster algorithm 
approximate complexity and entropy are trivially related via $c_\mu=1-h_\mu$. Thus,
there is no accentuated peak in the approximate complexity-entropy diagram.

%}}}1

\section{Summary}
\label{sect:summary}
In the presented article we performed numerical experiments to assess the performance
of three different entropy estimation algorithms that are based on symbolic substitution
methods. Binary test sequences where obtained by simulating the $2D$ Ising FM via 
single spin-flip dynamics at different temperatures $T$, thereby recording the orientation 
of a single spin.
We found that the most elaborate entropy estimation algorithm
yields results that are in good agreement with those obtained by an information
theoretic method based on the $M$-block Shannon entropy.
We further proposed and analyzed a measure that approximately accounts for the statistical
complexity of the binary sequences.  
The respective observable, termed \emph{approximate complexity}, can be understood in terms of
information theory. It measures the amount by which 
the entropy rate on the single-symbol level exceeds the asymptotic entropy rate and it
can easily be computed by means of black-box data-compression algorithms.
To support intuition on the gross behavior of the approximate complexity note that the larger
the correlations between the symbols in a given sequence, 
the larger the numerical value of the approximate complexity appears.
In the limits of completely ordered and fully random symbol sequences it assumes a value 
of zero. Therefore, the approximate complexity behaves
as one naively would expected for a quantity measuring the ``complexity''
of a system.
For all entropy estimation procedures considered, we find that the approximate complexity
is peaked at the critical point. Even for the less precise entropy estimation algorithms 
that systematically overestimate (underestimate) the entropy rate (approximate complexity), 
a finite-size scaling analysis in the sequence length shows that the peak of the approximate
complexity tends towards the critical point of the $2D$ Ising FM as the sequence length increases. 
Further, qualitative differences between the dynamics induced by a single-spin flip Metropolis
update and the Wolff cluster algorithm are discussed in terms of the information theoretic
observables.

For future work, we plan to apply these methods to systems
exhibiting quenched disorder, like spin glasses and random-field
systems, to find out whether the proposed methods will
work with similar high efficiency and precision.

\section{Acknowledgements}
OM acknowledges financial support from the DFG 
(\emph{Deutsche Forschungsgemeinschaft})
under grant HA3169/3-1.
The simulations were performed at the HERO cluster 
at the University of Oldenburg (Germany) which is funded  by the German Science
Foundation (DFG, INST 184/108-1 FUGG) and the
Minstry of Science and Culture (MWK) of the Lower Saxony state. 

\section*{References}

\bibliographystyle{unsrt}
\bibliography{lit_entropyComplexity.bib}

\begin{thebibliography}{10}

\bibitem{Shalizi2001}
C.~R. Shalizi and J.~P. Crutchfield.
\newblock {Computational Mechanics: Pattern and Prediction, Structure and
  Simplicity}.
\newblock {\em J. Stat. Phys.}, 104:817--878, 2001.

\bibitem{Loewenstern1995}
D.~Loewenstern, H.~Hirsh, P.~Yianilos, and M.~Noordewier.
\newblock {DNA Sequence Classification Using Compression-Based Induction}.
\newblock Technical Report DIMACS Tech. Rep. 95--04, DIMACS Center -- Rutgers
  University, 1977.

\bibitem{Ebeling1997}
W.~Ebeling.
\newblock {Prediction and entropy of nonlinear dynamical systems and symbolic
  sequences with LRO}.
\newblock {\em Physica D}, 109:42--52, 1997.

\bibitem{Baronchelli2005}
A.~Baronchelli, E.~Caglioti, and V.~Loreto.
\newblock {Measuring complexity with zippers}.
\newblock {\em Eur. J. Phys.}, page S69, 2005.

\bibitem{Grassberger2002}
P.~Grassberger.
\newblock {Data Compression and Entropy Estimates by Non-sequential Recursive
  Pair Substitution}.
\newblock 2002.

\bibitem{Puglisi2003}
A.~Puglisi, D.~Benedetto, E.~Caglioti, V.~Loreto, and A.~Vulpiani.
\newblock {Data compression and learning in time sequence analysis}.
\newblock {\em Physica D}, page~92, 2003.

\bibitem{crutchfield2010}
J.~P. Crutchfield and K.~Wiesner.
\newblock Simplicity and complexity.
\newblock {\em Physics World}, pages 36--38, February 2010.

\bibitem{cover2006}
T.~M. Cover and J.~A. Thomas.
\newblock {\em Elements of InformationTheory}.
\newblock Wiley, New York, 2006.

\bibitem{kolmogorov1963}
A.~N. Kolmogorov.
\newblock On tables of random numbers.
\newblock {\em Sankhya, the Indian Journal of Statistics A}, 25:369--376, 1963.

\bibitem{chaitin1987}
G.~Chaitin.
\newblock {\em Algorithmic Information Theory}.
\newblock Cambridge University Press, New York, 1987.

\bibitem{machta2006}
J.~Machta.
\newblock Complexity, parallel computation and statistical physics.
\newblock {\em Complexity}, 11:46--64, 2006.

\bibitem{Nagaraj2011}
N.~Nagaraj, M.~S. Kavalekalam, A.~Venugopal, and N.~Krishnan.
\newblock {Lossless Compression and Complexity of Chaotic Sequences}.
\newblock {\em (not published)}, 2011.
\newblock {A summary of this article is available at papercore.org, see
  {http://www.papercore.org/Nagaraj2011}}.

\bibitem{grassberger1986}
P.~Grassberger.
\newblock Toward a quantitative theory of self-generated complexity.
\newblock {\em Int. J. Theo. Phys.}, 25:907--938, 1986.

\bibitem{crutchfield1989}
J.~P. Crutchfield and K.~Wiesner.
\newblock Inferring statistical complexity.
\newblock {\em Phys. Rev. Lett.}, 63:105--108, 1989.

\bibitem{wiesner2011}
K.~Wiesner, M.~Gu, E.~Rieper, and V.~Vedral.
\newblock Information-theoretic bound on the energy cost of stochastic
  simulation, 2011.
\newblock preprint arXiv:1110.4217.

\bibitem{goldenfeld1992}
N.~Goldenfeld.
\newblock {\em {Lectures On Phase Transitions And The Renormalization Group}}.
\newblock Westview Press, Jackson, 1992.

\bibitem{Arnold1996}
D.~V. Arnold.
\newblock {Information-theoretic Analysis of Phase Transitions}.
\newblock {\em Complex Systems}, 10:143--155, 1996.

\bibitem{zlib}
{We used zlib version 1.2.3.3, see http://zlib.net}.

\bibitem{Ebeling1980}
W.~Ebeling and M.~A. Jimen\'ez-Monta\~no.
\newblock {On Grammars, Complexity, and Information Measures of Biological
  Macromolecules}.
\newblock {\em Math. Biosci.}, 52:53--71, 1980.

\bibitem{JimenezMontano2002}
M.~A. Jimenez-Montano, W.~Ebeling, and T.~Poeschel.
\newblock {SYNTAX: A computer program to compress a sequence and to estimate
  its information content}.
\newblock {\em (not published)}, 2002.

\bibitem{Benedetto2006}
D.~Benedetto, E.~Caglioti, and D.~Gabrielli.
\newblock {Non-sequential recursive pair substitution: some rigorous results}.
\newblock {\em J. Stat. Mech.}, page P09011, 2006.

\bibitem{Calcagnile2010}
L.~M. Calcagnile, S.~Galato, and Menconi G.
\newblock {Non-sequential Recursive Pair Substitutions and Numerical Entropy
  Estimates in Symbolic Dynamical Systems}.
\newblock {\em J. Nonlin. Sci.}, page 723, 2010.

\bibitem{Feldman2008}
D.~P. Feldman, C.~S. McTague, and J.~P. Crutchfield.
\newblock {The organization of intrinsic computation: Complexity-entropy
  diagrams and the diversity of natural information processing}.
\newblock {\em CHAOS}, 18:043106, 2008.

\bibitem{Ziv1977}
J.~Ziv and A.~Lempel.
\newblock {A Universal Algorithm for Sequential Data Compression}.
\newblock {\em IEEE Trans. Inform. Theory}, IT-23:337, 1977.

\bibitem{Bell1989}
T.~Bell, I.~H. Witten, and J.~G. Cleary.
\newblock {Modeling for Text Compression}.
\newblock {\em ACM Computing Surveys}, 21:557, 1989.

\bibitem{Crutchfield2003}
J.~P. Crutchfield and D.~P. Feldman.
\newblock {Regularities unseen, randomness observed: Levels of entropy
  convergence}.
\newblock {\em CHAOS}, 13:25--54, 2003.

\bibitem{papercore}
\emph{Papercore} is a free and open access database for summaries of scientific
  (currently mainly physics) papers, see {http://www.papercore.org/}.

\bibitem{Schuermann1996}
T.~Sch\"urmann and P.~Grassberger.
\newblock {Entropy estimation of symbol sequences}.
\newblock {\em CHAOS}, 6:414, 1996.

\bibitem{newman1999}
M.~E.~J. Newman and G.~T. Barkema.
\newblock {\em Monte {C}arlo Methods in Statistical Physics}.
\newblock Clarendon Press, Oxford, 1999.

\bibitem{Weiss2000}
O.~Weiss, M.~A. Jimen\'ez-Monta\~no, and H.~Herzel.
\newblock {Information Content of Protein Sequences}.
\newblock {\em J. theor. Biol.}, 206:379--386, 2000.

\bibitem{Wilms2011}
J.~Wilms, M.~Troyer, and F.~Verstraete.
\newblock Mutual information in classical spin models.
\newblock {\em J. Stat. Mech.}, 2011(10):P10011, 2011.

\bibitem{Erb2004}
I.~Erb and N.~Ay.
\newblock {Mulit-Information in the thermodynamic Limit}.
\newblock {\em J. Stat. Phys.}, 115:949, 2004.

\bibitem{JimenezMontano2002b}
M.~A Jimenez-Montano, W.~Ebeling, T.~Pohl, and P.~E. Rapp.
\newblock {Entropy and complexity of finite sequences as fluctuating
  quantities}.
\newblock {\em BioSystems}, 64:23--32, 2002.

\bibitem{practicalGuide2009}
A.~K. Hartmann.
\newblock {\em {Practical Guide to Computer Simulations}}.
\newblock World Scientific, Singapore, 2009.

\bibitem{wolff1989}
U.~Wolff.
\newblock {Collective Monte Carlo Updating for Spin Systems}.
\newblock {\em Phys. Rev. Lett.}, 62:361, 1989.

\end{thebibliography}

\end{document}